\documentclass[aps,prl,twocolumn, superscriptaddress,longbibliography]{revtex4-2}

\def\arc{$\alpha$-RuCl$_3$}

\usepackage{graphicx}
\usepackage{pstricks}
\usepackage{amsmath}
\usepackage{gensymb}
\usepackage{amsfonts}
\usepackage{dcolumn}
\usepackage{bm}
\usepackage{makecell}
\usepackage{xcolor}
\usepackage{float}
\usepackage{xcolor}
\usepackage[normalem]{ulem} 

\usepackage[colorlinks=True,linkcolor=red,citecolor=blue,urlcolor=blue]{hyperref}
\usepackage[capitalise]{cleveref}

\makeatletter
\AtBeginDocument{\let\LS@rot\@undefined}
\makeatother

\begin{document}
\title{{\it Ab initio} study of highly tunable charge transfer in $\beta$-RuCl$_3$/graphene heterostructures}
\author{Aleksandar Razpopov}
\email{razpopov@itp.uni-frankfurt.de}
\affiliation{Institut f\"ur Theoretische Physik, Goethe-Universit\"at Frankfurt, 60438 Frankfurt am Main, Germany}


\author{Roser Valent\'i}
\email{valenti@itp.uni-frankfurt.de}
\affiliation{Institut f\"ur Theoretische Physik, Goethe-Universit\"at Frankfurt, 60438 Frankfurt am Main, Germany}

\begin{abstract}

Heterostructures of graphene in proximity to magnetic insulators open the possibility to investigate exotic states emerging from the interplay of  magnetism, strain and charge transfer between the layers. Recent reports on the growth of self-integrated atomic wires of $\beta$-RuCl$_3$ on graphite suggest these materials as versatile candidates to investigate these effects. Here we present detailed first principles calculations on the charge transfer and electronic
structure of $\beta$-RuCl$_3$/heterostructures and provide a comparison with the work function analysis of the related honeycomb family members $\alpha$-RuX$_3$ (X = Cl,Br,I). We find that  proximity of the two layers leads to a hole-doped graphene and electron-doped RuX$_3$ in all cases,
which is sensitively dependent on the distance between the two layers. Furthermore, strain effects due to lattice mismatch control the magnetization  which itself has a strong effect on the charge transfer.
Charge accumulation in  $\beta$-RuCl$_3$ strongly drops away from the chain  making such heterostructures suitable candidates for sharp interfacial junctions in graphene-based devices.

 
\end{abstract}
\date{\today}
\maketitle



{\it Introduction.-} Graphene-based van der Waals (vdW) heterostructures reveal a variety of novel phenomena emerging from a notable modification of the electronic properties of their layer constituents due to proximity effects~\cite{Geim2013, Novoselov2016,yu2013vertically,britnell2012field,zhang2014ultrahigh,mehew2017fast,saraf2018emerging,ratha2013supercapacitor,kirubasankar2019sonochemical,sadeghi2016cross}.
Some examples are heterostructures of graphene with several  dichalcogenides like MoS$_2$, WS$_2$, and MoSe$_2$ with highly enhanced properties like ON/OFF current ratio in field effect transistors~\cite{yu2013vertically,britnell2012field}, photoresponsivity~\cite{zhang2014ultrahigh,mehew2017fast}, capacitance~\cite{saraf2018emerging,ratha2013supercapacitor,kirubasankar2019sonochemical}, or thermoelectricity~\cite{sadeghi2016cross}.
Such heterostructures are also desirable for possible technological applications as supercapacitors~\cite{saraf2018emerging}, photodetectors~\cite{aji2017high} and solid state lasers~\cite{zhao2015preparation}.

In recent years, heterostructures of graphene in proximity to magnetic insulators have gained increasing attention due to the possibility  to investigate new states resulting from the interplay between charge transfer, strain and magnetism in the heterostructures. For instance, Ref.~\cite{tseng2022gate} reported  hole-doping in graphene when placed in proximity to the magnetic insulators CrX$_3$ (X = Cl, Br, I), and found that the charge transfer from graphene to CrX$_3$ could be altered upon changing the magnetic states of the nearest magnetic insulator layer. Also, heterostructures of graphene with 
the Kitaev candidate material $\alpha$-RuCl$_3$ have been intensively investigated~\cite{zhou2019evidence, mashhadiArXiv19,Biswas_2019_electronic,rizzo2020,gerber2020abinitio,wang2020,balgley2022,Yang2022_magnetic,rossi2023direct}.
 These studies find a significant charge transfer from graphene to \arc\,, larger than in the CrI$_3$/gr (graphene) heterostructure,  giving rise to a van der Waals heterostructure that has higher conductivity than graphene. This charge transfer modifies the electronic properties
 of both materials and is influenced by the magnetism in $\alpha$-RuCl$_3$, as evidenced by the observation
 of anomalous quantum oscillations near the magnetic ordering temperature of \arc{},  that have 
  been interpreted within a Kitaev-Kondo lattice model~\cite{leeb2021}.

\begin{figure}[h!]
    \centering
    \includegraphics[width=\linewidth]{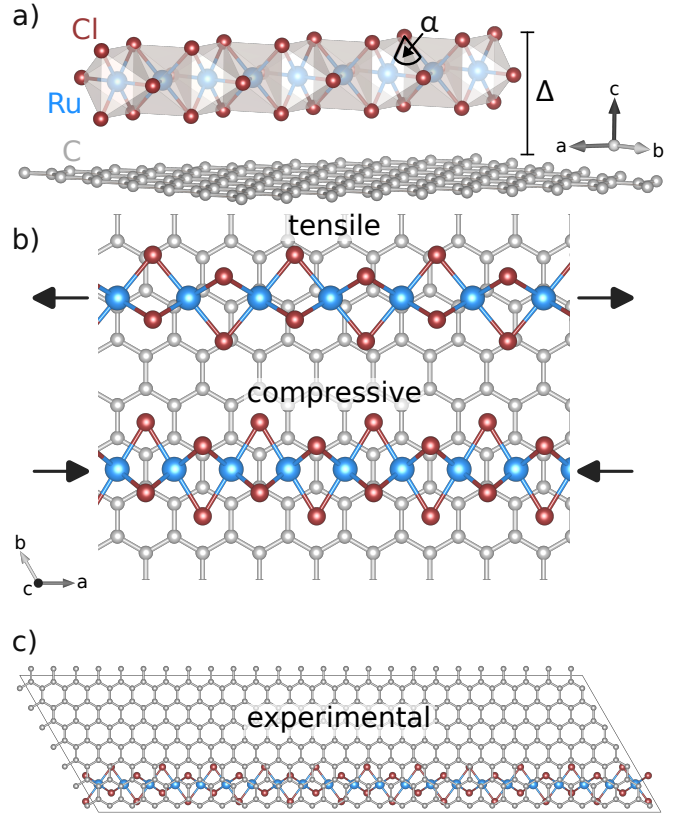}
    \caption{a) Crystal structure of the heterostructure $\beta$-RuCl$_3$/gr. $\beta$-RuCl$_3$ forms
    chains along the \textbf{a}-direction. Cl, Ru and C atoms are shown by dark red, light blue and grey balls, respectively. The chain height $\Delta$ is defined as the z coordinate of the highest Cl atom, and the angle $\alpha$ is between the Ru-Cl-Ru bond of the highest Cl atom. b) Cuts of $5\times5$ and $6\times6$ slabs of $\beta$-RuCl$_3$/gr considered in this study, which are commensurate with 7.8\% (tensile) and -13\% (compressive) strains, respectively,  as indicated by black arrows.  c) $7\times25$ slab 
    of  $\beta$-RuCl$_3$/gr emulating the experimentally reported strained heterostructres~\cite{tomoya2023_growth}. }
    \label{fig:Crystal_struct}
\end{figure}

  Motivated by this versatility of phases in heterostructures of graphene in proximity to magnetic insulators
  and in the search for systems where charge transfer could be considerable enough to modify the electronic properties of both layers,
  we explore here the charge transfer of heterostructures of graphene with  RuBr$_3$~\cite{imai2022zigzag} and RuI$_3$~\cite{nawa2021strongly,ni2022honeycomb},  which are
  isostructural honeycomb van der Waals materials to the Kitaev candidate material \arc{}~\cite{winter2017models,motome2020hunting,kaib2022electronic,kim2022alpha,trebst2022kitaev,ahn2024progress}, and we present detailed
  calculations of the heterostructure $\beta$-RuCl$_3$/gr, where the chain-based $\beta$-RuCl$_3$ has recently been shown to grow as atomic-scale wires on graphite~\cite{tomoya2023_growth}.

 We proceed by first calculating the work functions for the various compounds within {\it ab initio} density functional theory (DFT). Relative comparisons to the
   work function of graphene provide an indication of the amount of expected charge transfer in the
   heterostructures. We then perform detailed calculations for 
   $\beta$-RuCl$_3$/gr by considering various strained structures adapted to the lattice mismatch between graphene and the magnetic insulator.
   Our main observations are that proximity of the two layers leads to a hole-doped graphene and electron-doped RuX$_3$ in all cases,
   which is sensitively dependent on the distance between the two layers.
   We further find that strain effects due to lattice mismatch, in first order, control the magnetization  which itself has a strong effect on the charge transfer. Moreover, charge accumulation in  $\beta$-RuCl$_3$ sharply drops away from the chain.

{\it Methods.-} 
In order to obtain the work functions of the various systems and to calculate the charge transfer and electronic properties of the heterostructure $\beta$-RuCl$_3$/gr  we use the projector augmented wave method~\cite{Bloechl1994} as implemented in the VASP simulation package \cite{Kresse} version 6.3.0.
As exchange-correlation functional within DFT, we consider the generalized gradient approximation (GGA) \cite{Perdew1996}  including the Coulomb corrections on the Ru 4d orbitals using the GGA+U scheme~\cite{Dudarev}.
Additionally, van der Waals corrections are included via the DFT+D2 method of Grimme~\cite{Grimme}.
For structural relaxations we consider a $6\times6\times1$ k-mesh, while for electronic property calculations
we consider a $12\times12\times1$ k-mesh that provides accurate results.

For the work function estimates we keep the crystal structures as provided by X-ray data~\cite{x-ray_fletcher_1967} without further relaxing them. 
The work function $W$ is given by~\cite{kahn2016fermi} 
\begin{equation}
    W = E_{\rm vac} - E_{\rm F},
\end{equation}
where $E_{\rm vac}$ is the vacuum energy and $E_{\rm F}$ is the Fermi energy.
The Fermi level is placed at the middle of the insulating gap.
Note that for these calculations only monolayers of the corresponding materials were considered since previous DFT studies showed that the work function is primarily related to the surface properties~\cite{Kim2021thickness}.

The heterostructure $\beta$-RuCl$_3$/gr is constructed from the experimental 
crystal structure of $\beta$-RuCl$_3$~\cite{x-ray_fletcher_1967} as shown in Fig.~\ref{fig:Crystal_struct} a) top panel.
For the slab calculations a void of $d_{\rm layer} = 20$~\AA~ is considered to ensure that the heterostructure is electronically isolated from the neighboring ones along the c-direction
since the calculations assume periodic boundary conditions. As the graphene layer has higher stiffness than $\beta$-RuCl$_3$, its structure is fixed with
$d_{\rm C-C}$ = 1.420~\AA\
and only $\beta$-RuCl$_3$ is relaxed.
The charge transfer is obtained via Bader analysis~\cite{HENKELMAN_2006_a_fast} of the optimized structure.

{\it Results.-} In Table~\ref{tab:work_function} we present the calculated work functions 
$W_{\rm DFT}$ for the systems considered in this study.
For $\beta$-RuCl$_3$ we also include the stress dependence of the work functions. 
For graphene and $\alpha$-RuCl$_3$ our calculated values are in excellent agreement with the $W_{\exp}$ values reported in earlier experimental works~\cite{Rutkov_2020_graphene,pollini_1996_electronic,Klaproth_2022_work_fuction}.
\begin{table}[h]
\centering
\begin{tabular}{c|c|c}
Materials & $W_{\rm DFT}$[eV] & $W_{\exp}$[eV] \\ \hline \hline
graphene  & 4.20 & 4.30~\cite{Rutkov_2020_graphene} \\ 
$\beta$-RuCl$_3$  (-13\%)& 6.26 & - \\
$\beta$-RuCl$_3$       &  6.04 & - \\
$\beta$-RuCl$_3$  (7.8\%)& 5.97 & -\\
$\alpha$-RuCl$_3$  & 6.01 &  6.10~\cite{pollini_1996_electronic,Klaproth_2022_work_fuction} \\
RuBr$_3$           & 5.64 & -\\
RuI$_3$            & 5.23 & -
\end{tabular}
\caption{Calculated work functions of graphene, and monolayers of RuX$_3$ (X = Cl, Br, I)  and $\beta$-RuCl$_3$, under various stress conditions for the latter
along the chain direction (compressive (-13\%) and tensile (7.8\%), compare with Fig.~\ref{fig:Crystal_struct}). The calculated value for graphene agree with previous reported results including spin-orbit coupling (SOC) and considering  HSE06 as exchange-correlation functional~\cite{zhang2018strong}. }
\label{tab:work_function}
\end{table}
The differences in work functions between graphene and the various RuX$_3$ monolayers already suggest that in all cases a graphene hole-doping is to be expected
in the RuX$_3$/gr heretostructures. The charge transfer is expected to increase in going from RuI$_3$ to RuCl$_3$
Interestingly, while the work functions for $\alpha$- and $\beta$-RuCl$_3$ are similar,  the effect of strain  has an important impact on the resulting work function of
$\beta$-RuCl$_3$ which increases under compressive strain and decreases under tensile strain. This effect is less remarkable in  $\alpha$-RuCl$_3$~\cite{Biswas_2019_electronic}.
The above results suggest that $\beta$-RuCl$_3$/gr under various conditions of strain
may be a promising candidate to achieve considerable charge transfer, and therefore we concentrate in what follows on analyzing its electronic properties.

In order to simulate a strained heterostructure one needs to consider slabs that are commensurate with the chosen strain. The smaller the strain values, the
larger the corresponding slabs that need to be considered. Balancing between  numerical feasibility and complexity, we start with   $5\times5$  and  $6\times6$ slabs  which allow to describe  $\beta$-RuCl$_3$/gr 
with  $\beta$-RuCl$_3$ under  7.8\% (tensile)  and -13.0\% (compressive) strain, respectively (see Fig.~\ref{fig:Crystal_struct} b)).
We first obtain geometrically optimized slab structures of $\beta$-RuCl$_3$/gr  (see SI for details) for these two extreme strained cases, and we then consider slab structures that are commensurate  with the strain achieved in recent experiments~\cite{tomoya2023_growth,private_tomoya}  (compressive, (-1.8\%)) (Fig.~\ref{fig:Crystal_struct} c)). All the reported values of strain are measured with respect to bulk $\beta$-RuCl$_3$ lattice parameters.
Two different initial magnetic configurations of Ru chains were also considered for the relaxation of the slabs: ferromagnetic (FM) and  antiferromagnetic (AFM)  (see Fig. S2 in SI).
In a second step we analyze the electronic properties and charge transfer for the various slabs.

Starting with the  7.8\% tensile configuration,
the initial geometry of $\beta$-RuCl$_3$ consists of a uniform Ru-Ru chain, where the Ru atoms are centered above a C atom and in-between two C atoms of graphene in an alternating fashion
(Fig.~\ref{fig:Crystal_struct} b)). 
After the structural relaxation within GGA+U, $U-J= U_{\rm eff}=3.0$ eV and a FM configuration of Ru, the final configuration switches from FM to double-AFM (dAFM), with two-up and two-down magnetic moments along the Ru chain (see  Fig. S2 in SI).
This relaxed structure displays three different Ru-Ru distances:  $d_{\rm Ru-Ru}^{(1)}$ = 3.138~\AA,  $d_{\rm Ru-Ru}^{(2)}$ = 3.065~\AA, and $d_{\rm Ru-Ru}^{(3)}$ = 2.949~\AA. 
The amount of dimerization of the chain, given by the ratio of the longest to shortest Ru-Ru bond-distance, is $\approx$ 6\%.
Note that this configuration shows a weaker chain  dimerization than the reported dimerized structure of bulk $\beta$-RuCl$_3$ of $\approx$ 13\%  at low temperatures~\cite{hillebrecht2004_about}.
The associated Ru-Cl-Ru angles along the chain $\alpha$ (Fig.~\ref{fig:Crystal_struct} a)) get modified as well, introducing a slight non-linearity in the Ru chains with the angles lying between  $0.3^{\circ} - 0.7^{\circ}$ compared to a perfectly linear chain of Ru atoms.  We also performed structural relaxations  starting with an AFM configutation
and we reached the same optimized structure and magnetic configuration as described above, showing that the final relaxed geometry is not dependent on the initial magnetic configuration.

Total energy calculations of the optimized structures show that FM and dAFM configurations
are energetically highly competing states. 
For further analysis of the optimized structure, as shown below,  we will  consider therefore, for simplicity, 
the optimized strained structure with Ru in a FM configuration.

\begin{figure}[h!]
    \centering
    \includegraphics[width=\linewidth]{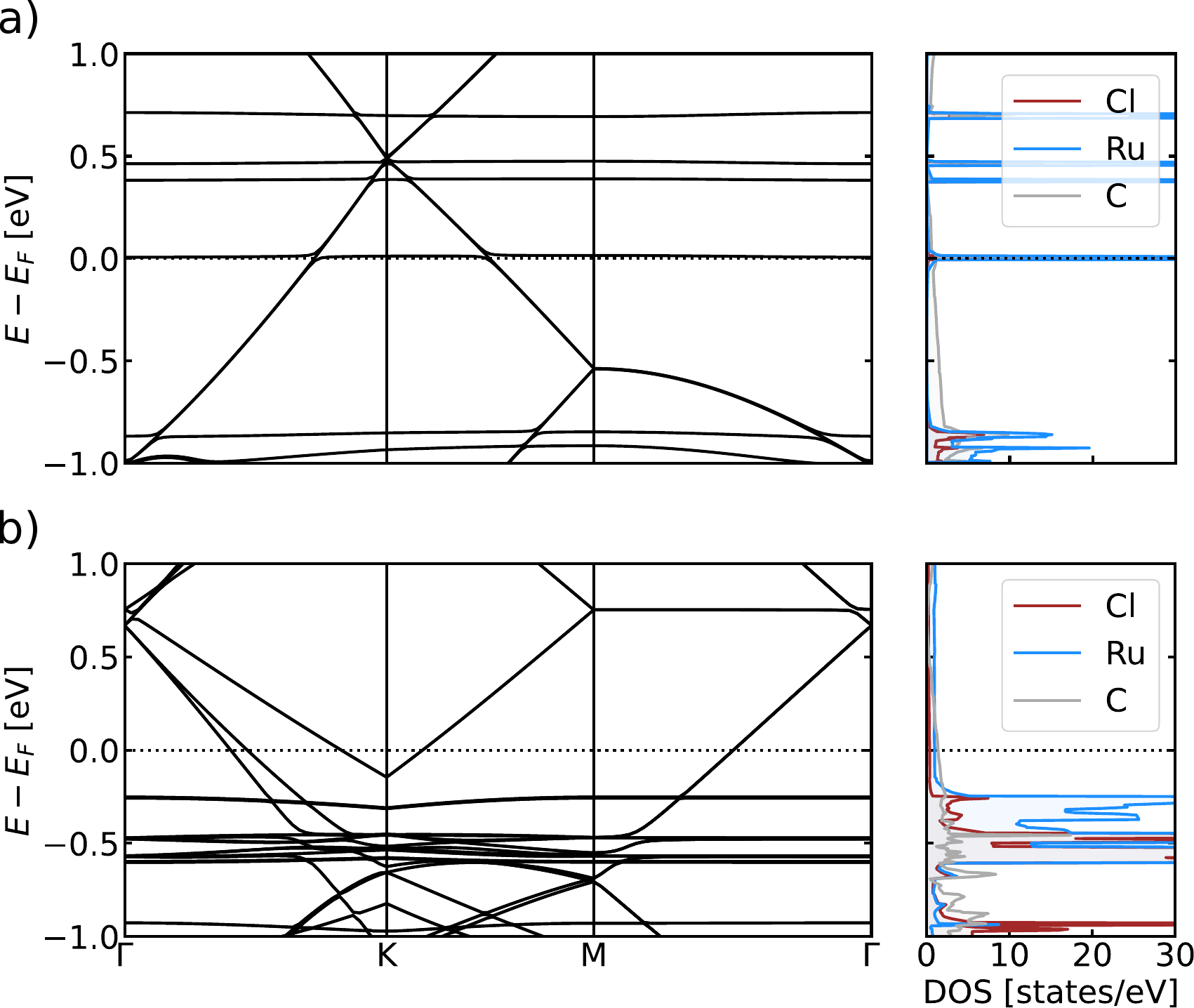}
    \caption{Band structure  and atom-resolved DOS of a) $\beta$-RuCl$_3$/gr under 7.8\%
    tensile strain with FM ordered moments on the Ru sites. The Dirac point of pristine graphene at the \textbf{K} point is shifted $\approx$ 0.50 eV above the Fermi level. b) $\beta$-RuCl$_3$/gr under -13\%
    compressive strain. The Dirac point of graphene at the \boldsymbol{$\Gamma$} point is shifted by $\approx$ 0.65 eV above the Fermi surface. In both cases there is charge transfer from graphene to the $\beta$-RuCl$_3$ chain. The results were obtained with  GGA+U, with $U-J= U_{\rm eff}=3.0$ eV.}
    \label{fig:electronic}
\end{figure}

\cref{fig:electronic} {a)} displays the calculated  electronic band structure and atom resolved density of states (DOS) within GGA+U for 
the 7.8\%  strained $\beta$-RuCl$_3$/gr heterostructure. Note that the $5\times5$-graphene's \textbf{K} point folds back on itself~\cite{zhou2013_3n}.
We observe that the Dirac point shifts by about 0.50 eV above the Fermi energy $E_F$ compared to pure graphene, indicating hole doping (charge depletion) in the graphene layer. This shift of the Dirac point is consistent with a positive value of the calculated difference in work functions ($W_{\rm RuCl_3} - W_{\rm Graphene}\approx$ 2 eV).
 
In contrast to the Mott insulating behavior of bulk  $\beta$-RuCl$_3$~\cite{tomoya2023_growth}, the heterostructure $\beta$-RuCl$_3$/gr undergoes an insulator to metal transition due to the chemical shift. The electronic weights at the Fermi level are mostly of  Ru $d$ orbitals hybridizing with Cl and C $p$ orbitals (see DOS in \cref{fig:electronic} a)). The Ru flat bands pinned near the Fermi level
are a consequence of the tensile strain on $\beta$-RuCl$_3$  which
 drives the chain towards an atomic limit with  localized bands.

Using Bader analysis~\cite{HENKELMAN_2006_a_fast} we compute the charge transfer per formula unit between $\beta$-RuCl$_3$ and graphene.
 $\beta$-RuCl$_3$ is electron-doped with $\delta \rho_{\rm RuCl_3}  = 0.085$ e/f.u. and
 graphene hole-doped with $\delta \rho_{\rm C}  = -0.007$ e/C.
These values have been also confirmed for AFM and dAFM configurations of Ru, where we observed negligible differences of less than 5\% due to numerical approximations. 
We also studied the dependence of the charge transfer on the presence of Ru magnetic moments by performing non-magnetic calculations and found a significant increase of the charge transfer by $\approx$ 37\%, compared to the magnetic case.
This result indicates a  strong correlation between  magnetism and charge transfer in this system.

Next, we investigate $\beta$-RuCl$_3$/gr under compressive strain (\cref{fig:electronic} b)) by 
considering a $6\times6$ slab  where initially Ru atoms are positioned at the center of the carbon honeycomb in graphene. Structural relaxation with both,  FM and AFM initial magnetic configurations for Ru  lead to the same relaxed non-magnetic solution where d$_{\rm Ru-Ru}$ = 2.459~\AA~  forming  regular linear chains (Fig.~\ref{fig:Crystal_struct} b)), in contrast to the tensile heterostructure above.
The reduced Ru-Ru distance corresponds to 13\% compressive strain compared to the bulk undimerized structure with $d_{\rm Ru-Ru}^{\rm Bulk}$ = 2.861~\AA.
The height of the $\beta$-RuCl$_3$ to graphene, $\Delta$ (\cref{fig:Crystal_struct} a)), increases from 6.5~\AA~to 6.7~\AA~going from tensile to compressive strain as a consequence of the shortening of the Ru-Ru distance that increases the angle~$\alpha$ changing the value of the height.
This height profile of the relaxed heterostructures is close to the obtained from experiment~\cite{tomoya2023_growth}.

In \cref{fig:electronic} b) we show the
calculated band structure and DOS for this case. In contrast to the tensile case, no Ru states are pinned near the Fermi level. The reason for that is the short Ru-Ru distance due to the compressed chain structure
inducing Ru-Ru bonding-antibonding band splittings located at -0.5eV and 2.5eV respectively (see Fig. S6 in SI). 
The Dirac point of pristine graphene is folded back to the \boldsymbol{$\Gamma$} point of the heterostructure due to the choice of the super-cell~\cite{zhou2013_3n} and shifted $\approx$ 0.65 eV above the Fermi energy.
This indicates a higher charge transfer compared to the tensile strain. This result is in agreement to the trends found in the  $W_{\rm DFT}$ differences (Table~\ref{tab:work_function}), that suggest an increase of charge transfer from tensile to compressive strain. 
\begin{figure}
    \centering
    \includegraphics[width=\linewidth]{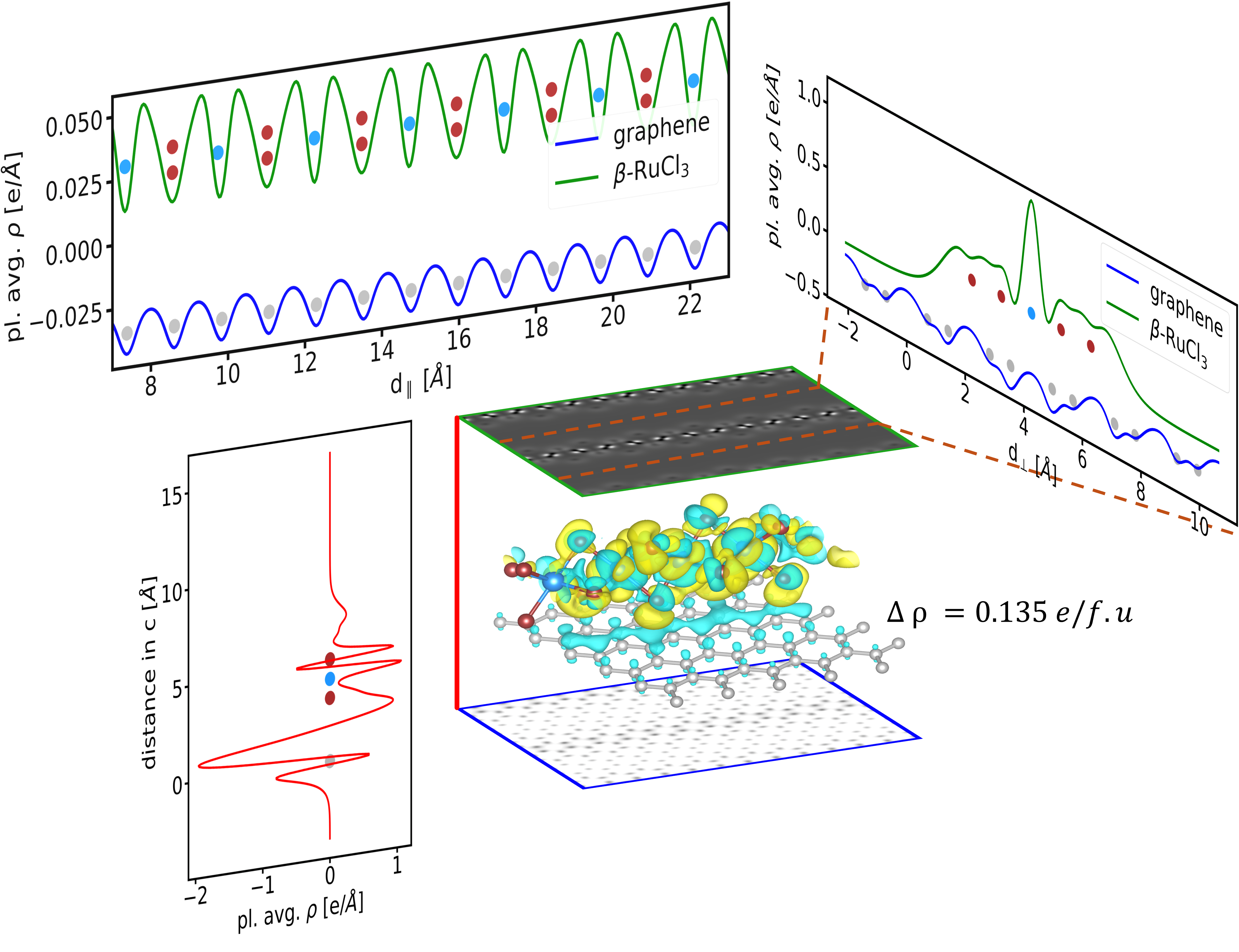}
    \caption{Spatial charge redistribution in the (-13\%) compressive strained $\beta$-RuCl$_3$/gr, $\Delta \rho = \Delta \rho_{\rm RuCl_3/gr} - \Delta \rho_{\rm RuCl_3} - \Delta \rho_{\rm gr}$. The figure shows in the top panels
    a two-dimensional projection of the planer-average charge density (pl. avg. $\Delta \rho$)  along the parallel ($d_{\parallel}$) and perpendicular ($d_{\perp}$) chain directions in the $ab$ plane. Dots in grey, light blue and dark red  symbolize the position of the C, Ru and Cl atoms. The green and blue curves
    display the 2D projection of pl. avg. $\Delta \rho$  within the $\beta$-RuCl$_3$ and graphene layers, respectively.  The central panel shows a three-dimensional view of   $\Delta \rho$. In the bottom left panel the total charge redistribution along $c$ is shown as a red curve. The planer-average charge values are multiplied by a factor of 1000.}
    \label{fig:charge_destribution}
\end{figure}


The Bader analysis~\cite{HENKELMAN_2006_a_fast}~ for the compressed  $\beta$-RuCl$_3$/gr confirms that the charge transfer is strongly enhanced compared to the tensile strained
heterostructure.
We observe electron  doping in  $\beta$-RuCl$_3$, $\delta \rho_{\rm RuCl_3}  = 0.134$ e/f.u., which is 37\%~larger than in the tensile case, and hole doping in graphene with $\delta \rho_{\rm C}  = -0.012$~e/C.
This amount of charge transfer coincides with the non-magnetic estimates for the tensile strained heterostructure and corroborates the observation of a strong correlation between the absolute value of the Ru magnetic moment and the charge transfer.

 \cref{fig:charge_destribution} displays the spatial charge redistribution along the three directions for the  compressive strained $\beta$-RuCl$_3$/gr (results for the tensile strained heterostructure are shown in SI). 
Along the c-direction we observe a strong charge depletion (negative $\Delta \rho$) at the graphene layer and charge accumulation (positive $\Delta \rho$)
at the $\beta$-RuCl$_3$ layer (\cref{fig:charge_destribution} bottom left panel).
Along the chain direction ($d_\parallel$,  \cref{fig:charge_destribution} top left panel) the charge distribution follows a uniform pattern, while
in the direction perpendicular to it ($d_\perp$,  \cref{fig:charge_destribution} top right panel), the charge accumulation at the atomic wire sharply decays and goes to zero within a few Angstroms away from the chain, similar to what has been observed in $\alpha$-RuCl$_3$ stripes on graphene~\cite{balgley2022}. 
We further investigated the dependence of the relative orientation of the $\beta$-RuCl$_3$ chains to graphene keeping the height constant, but
we didn't find any significant changes in the charge transfer (see SI).

In order to estimate whether the above considered heterostructures are realizable we calculate the formation enthalpy  $\Delta H$ defined as 
\begin{equation}
    \Delta H = E(AB) - E(A) - E(B),
\end{equation}
where $E(AB)$ is the energy of the $\beta$-RuCl$_3$/gr heterostructure, and $E(A)$ and $E(B)$ are the energies of the subsystems A (graphene) and B ($\beta$-RuCl$_3$-chain).
For the tensile strained $\beta$-RuCl$_3$/gr,   $\Delta H_T$ = -0.230 meV/f.u and for the compressive strained $\beta$-RuCl$_3$/gr,  $\Delta H_C$ = -0.320 meV/f.u.
From this analysis we conclude that both strained heterostructures are stable and could be realized in experiments.

We proceed now with the study of a slab of $\beta$-RuCl$_3$/gr  (see Fig.~\ref{fig:Crystal_struct} c)) corresponding to recently reported strained heterostructures~\cite{tomoya2023_growth,private_tomoya} 
showing $\approx$ 2\% compressive strain.
Since the analysis performed so far in the $5\times5$ and $6\times6$ slabs suggests no significant dependence on the relative orientation of the chains and graphene,
 we do not  relax the full heterostructure, what would lead to unsustainable heavy calculations due to the large number of atoms in the slab. 
Instead, we select for the height of the chain $\Delta$ (Fig.~\ref{fig:Crystal_struct} a)) a value close to the estimates obtained
for the above analyzed heterostructures and we consider Ru magnetic moments initially in a FM configuration. We find via Bader analysis a charge transfer of $\delta \rho_{\rm RuCl_3}  = 0.113$ e/f.u.
and  $\delta \rho_{\rm C}  = -0.007$~e/C.  To compare the results of the three heterostructures we plot
in
\begin{figure}[h]
    \centering
    \includegraphics[width=\linewidth]{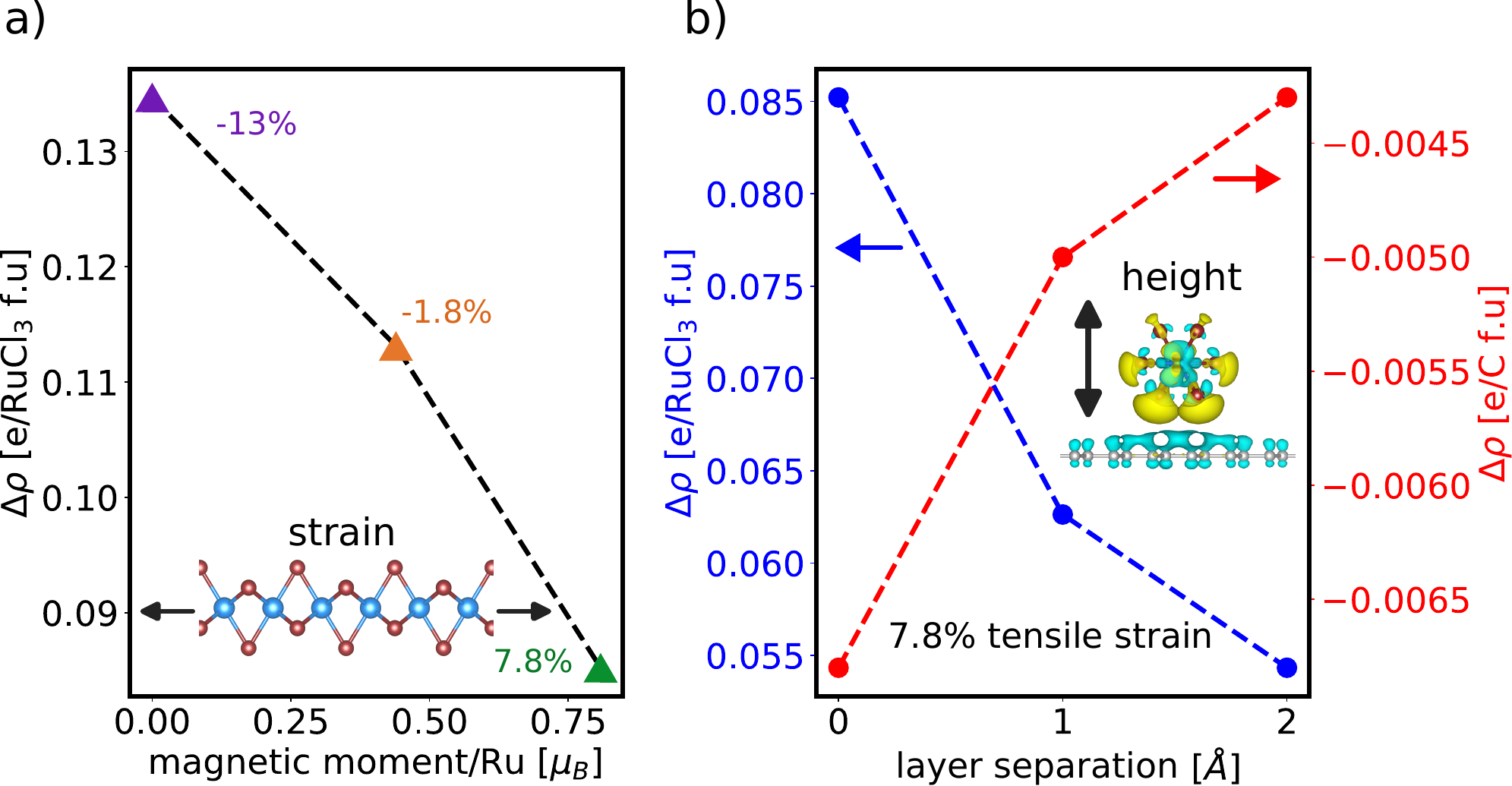}
    \caption{Possible charge-transfer tuning parameters. a) Dependence between the magnetic moment and the charge transfer. Each colored data point refers to a different heretostructure setting: purple to tensile strain, orange to experimental strain and green to compressive strain. The strain symbolizes a way to control the magnetic moment. b) Charge transfer in the tensile strained structure as a function of the manually controlled layer separation distance.}
    \label{fig:charge-transfer-zero-strain}
\end{figure}
  \cref{fig:charge-transfer-zero-strain} a)  the dependence of the charge transfer on the average magnetic moment on the Ru sites.  We observe a direct relation between a decrease of charge with an increase of magnetic moment. Actually, the value of the
charge transfer for all three cases in  non-magnetic calculations is the same. These results suggest that strain in first order controls the magnetization which itself has a strong effect on the charge transfer. 
Besides strain, changing the height $\Delta$  offers as well a possibility to modify the charge transfer. Increasing the layer separation between $\beta$-RuCl$_3$ and graphene leads to decreasing charge transfer. In 
\cref{fig:charge-transfer-zero-strain} b) we show this dependence for the tensile heterostructure.  This behavior is to be
expected and has been observed in other van der Waals heterostructures~\cite{crippa2024heavy}. 
Experimentally, it is also possible to control the distance between two atomic wires as shown by the synthesis technique of $\beta$-RuCl$_3$ used in Ref.~\cite{tomoya2023_growth}. We therefore analyzed the charge transfer as a function of interchain distance but no significant changes  were observed
as can be already  preempted by the fact that charge accumulation  perpendicular to the $\beta$-RuCl$_3$ chains sharply drops at the edges (see \cref{fig:charge_destribution} top right panel).


%

{\it Conclusions.-} We have investigated via first principles calculations,
 the charge transfer and electronic properties of heterostructures of graphene with  the honeycomb-based $\alpha$-RuX$_3$, (X = Cl, Br, I)  and chain-based
$\beta$-RuCl$_3$ by calculating work functions and by performing slab simulations of the recently reported $\beta$-RuCl$_3$/gr heterostructure~\cite{tomoya2023_growth}.
Our calculations on three exemplary heterostructures of $\beta$-RuCl$_3$/gr including different degrees of tensile and compressive strain 
showed that proximity of the two layers leads to a hole-doped graphene and electron-doped RuX$_3$ in all cases,
   which is sensitively dependent on the distance between the two layers. Furthermore,
strain effects due to lattice mismatch, control the magnetization  which itself has a strong effect on the charge transfer, and
charge accumulation in  $\beta$-RuCl$_3$ sharply drops away from the chain what opens the
possibility for device development.
%
%

{\it Acknowledgements}.- The authors are grateful to S. Biswas, T. Asaba, K. Burch, and E. Henriksen for valuable discussions and gratefully acknowledge support by the Deutsche Forschungsgemeinschaft (DFG, German Research Foundation) for funding through Project No. 509751747 (VA 117/23-1) and TRR 288 — 422213477 (project A05). 
\clearpage
\widetext
\appendix
\begin{center}
		\textbf{\large \textsc{Supplementary Information}:\\ {\it Ab initio} study of highly tunable charge transfer in $\beta$-RuCl$_3$/graphene heterostructures} \bigskip \bigskip
	\end{center}
	\twocolumngrid
	
	\setcounter{equation}{0}
	\setcounter{figure}{0}
	\setcounter{table}{0}
	\setcounter{page}{1}
	
	\makeatletter
	\renewcommand{\theequation}{S\arabic{equation}}
	\renewcommand{\figurename}{Supplementary Figure}
	\setcounter{figure}{0}  
	\renewcommand{\thetable}{S\Roman{table}}
	
	\DeclareFontShape{OT1}{cmss}{m}{it}{<->ssub*cmss/m/sl}{}
	\renewcommand{\rmdefault}{cmss}
	\renewcommand{\sfdefault}{cmss}
	
	\DeclareFontFamily{OT1}{cmbr}{\hyphenchar\font45 }
	\DeclareFontShape{OT1}{cmbr}{m}{n}{%
		<-9>cmbr8
		<9-10>cmbr9
		<10-17>cmbr10
		<17->cmbr17
	}{}
	\DeclareFontShape{OT1}{cmbr}{m}{sl}{%
		<-9>cmbrsl8
		<9-10>cmbrsl9
		<10-17>cmbrsl10
		<17->cmbrsl17
	}{}
	\DeclareFontShape{OT1}{cmbr}{m}{it}{%
		<->ssub*cmbr/m/sl
	}{}
	\DeclareFontShape{OT1}{cmbr}{b}{n}{%
		<->ssub*cmbr/bx/n
	}{}
	\DeclareFontShape{OT1}{cmbr}{bx}{n}{%
		<->cmbrbx10
	}{}

	
\author{Aleksandar Razpopov}
\email{razpopov@itp.uni-frankfurt.de}
\affiliation{Institut f\"ur Theoretische Physik, Goethe-Universit\"at Frankfurt, 60438 Frankfurt am Main, Germany}

\author{Roser Valent\'i}
\email{valenti@itp.uni-frankfurt.de}
\affiliation{Institut f\"ur Theoretische Physik, Goethe-Universit\"at Frankfurt, 60438 Frankfurt am Main, Germany}

\subsection*{Supplementary Note 1: Work function}
We calculate via ab initio density functional theory (DFT) the work function $W$ for graphene and the RuX$_3$ family with X=Cl, Br, I and compare with the experimental values if they are available (see main text), as implemented in VASP~\cite{Kresse}.
The simulations are performed on a  10$\times$10$\times$5 $\Gamma$-centered \textbf{k}-mesh for $\alpha$-RuCl$_3$, RuBr$_3$, and RuI$_3$ and for $\beta$-RuCl$_3$ on a 8$\times$1$\times$22.
As exchange correlation functional we use GGA~\cite{Perdew1996} and include the Coulomb corrections on the Ru 4d orbitals using the DFT+U scheme~\cite{Dudarev}.
We set $U_{\rm eff}=~3.0,~2.5, ~2.0,~\text{and}~1.5$~eV for $\beta$-RuCl$_3$, $\alpha$-RuCl$_3$, RuBr$_3$ and RuI$_3$ motivated by constrained random-phase approximation (cRPA) estimates~\cite{kaib2022electronic}, respectively.
Spin orbit effects are taken into account.
The basis set plane wave cut-off for the expansion is set to 600~eV in each calculation.
For all systems the work function $W$ has been checked with respect to the size of the void, where all values are given for a void of the size of 2$\times$\textbf{c} lattice constant of the responding material.

\subsection*{Supplementary Note 2: Heterostructures calculations}
Using DFT we optimize the heterostructure of $\beta$-RuCl$_3$/gr for tensile (+7.8\%) and compressive strain (-13\%), where we keep the graphene layer fixed, as graphene has high stiffness. 
The tensile strain simulation is performed with a 2$\times$1-$\beta$-RuCl$_3$ chain on 5$\times$5 graphene layer.
In the initial geometry the Ru atoms are centered alternating above a C atom and in between a C-C bond.
The compressive strain heterostruture consists of a 3$\times$1 $\beta$-RuCl$_3$ chain on 6$\times$6 graphene, where each Ru atom is initially set above the center of a honeycomb built by the C atoms.
The initial geometry of each structure consists of uniform chain, the simulated unit cells are shown in Supplementary~\cref{fig:heterostructure}.
The experimental strain (not relaxed) consists of 11$\times$1 $\beta$-RuCl$_3$ chain on 7$\times$25 graphene.
We include the Coulomb corrections on the Ru 4d orbitals using the DFT+U scheme~\cite{Dudarev}, with $U_{\rm eff}=~3.0$~eV.
The calculations are carried on a 6$\times$6$\times$1 \textbf{k}-grid for the compressive and tensile case, and on a 1$\times$2$\times$1 \textbf{k}-grid for the experimental strain with a basis set plane wave cut-off for the expansion is set to 600~eV.
The graphene layer distance is set to $d_{\rm layer} = 20$~\AA~with C-C distance within the graphene layer of $d_{\rm C-C}$ = 1.420~\AA.
This large layer distance $d_{\rm layer}$ ensures that the layer-layer interactions can be neglected and considered as a 2D system. 
The  heterostructure is relaxed until the forces decrease down to $10^{-3}$~eV/\AA~and energy convergence criterion of $1\times10^{-6}$ eV. 

\begin{figure}[h]
    \centering
    \includegraphics[width=\linewidth]{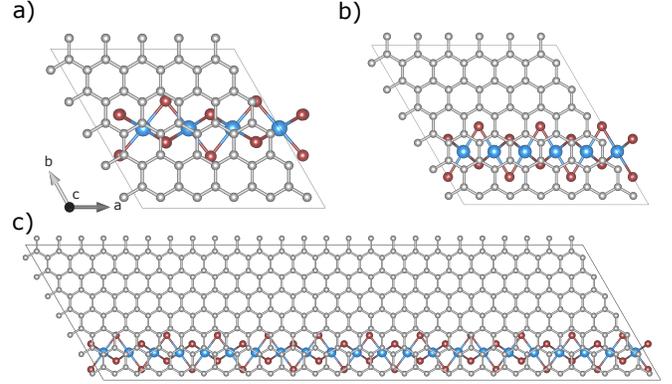}
    \caption{In panel \textbf{a)} we show the unit cell size used for the tensile strain, \textbf{b)} compressive unit cell size. In \textbf{c)} we show the unit cell used to simulate the experimental strain.}
    \label{fig:heterostructure}
\end{figure}

For each strain the structures are relaxed using ferromagnetic (FM), antiferromagnetic (AFM), and double-antiferromagnetic (dAFM) ordered magnetic, symbolized in Supplementary~\cref{fig:magnetic-order}.
We find that final structure does not depend on the magnetic order.
Starting from the AFM state we obtain $d_{\rm Ru-Ru}^{(1)}$ = 3.137~\AA, $d_{\rm Ru-Ru}^{(2)}$ =  3.063~\AA, and  $d_{\rm Ru-Ru}^{(3)}$ = 2.951~\AA.
These values show very small deviation compared to the FM and dAFM state (see main text). 

The electronic properties of the system are calculated in each relaxed structure with a more densed \textbf{k}-grid of 12$\times$12$\times$1.
When comparing the total energy of each magnetic configuration, we find that FM state is energetically most favorable state, with 4 meV/f.u. lower compared to the dAFM order.
The AFM state is the energetic highest state with 84 meV/f.u. above the dAFM solution.

\begin{figure}[h]
    \centering
    \includegraphics[width=\linewidth]{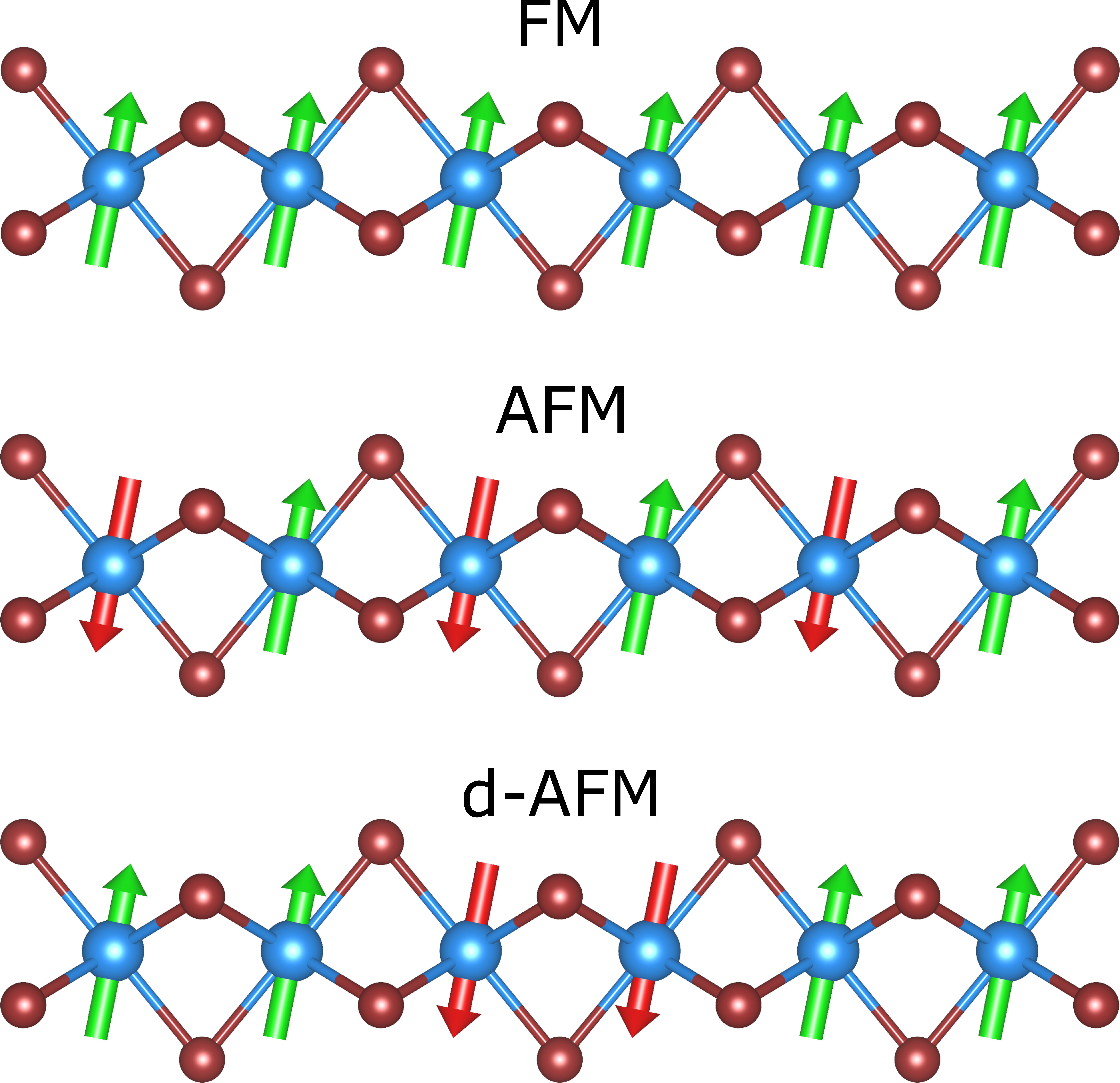}
    \caption{Different magnetic orders investigated during this study: ferromagnetic order (FM), antiferromagnetic (AFM), and double antiferromagnetic (d-AFM). The green up arrows symbolize the magnetic moment with spin up, and the red arrows the spin down.}
    \label{fig:magnetic-order}
\end{figure}

\subsection*{Supplementary Note 3: Geometry dependence}

We study the dependence of the charge transfer and the relative orientation between the $\beta$-RuCl$_3$ atomic wire and the graphene layer.
For this we considered four additional chain orientations in the compressed case, see Supplementary~\cref{fig:additional_settings_B}.
Here, we keep the height $\Delta$ from the first relaxed structure and shift manually the chain in the \textbf{a-b} plane.
Via Bader analysis~\cite{HENKELMAN_2006_a_fast} we computed the charge transfer for each case, the values in the $\beta$-RuCl$_3$ layer are noted in Supplementary~\cref{fig:additional_settings_B}.
We observe negligible difference in the charge transfer. 

\begin{figure}[h]
    \centering
    \includegraphics[width=\linewidth]{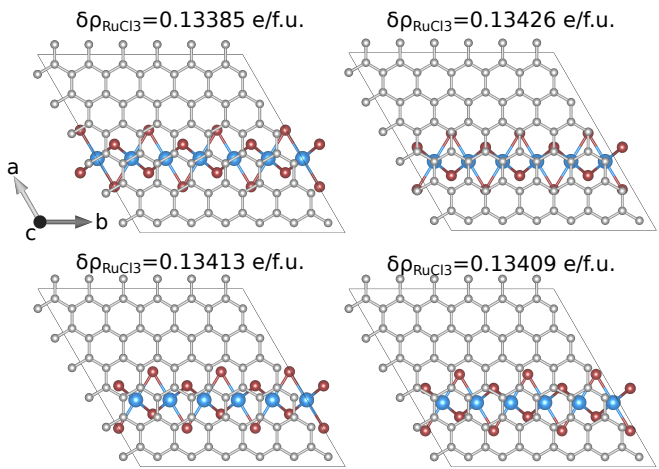}
    \caption{Additional relative geometries that have been investigated in the compressive setting. Above each setting we note the charge transfer $\delta \rho_{\rm RuCl_3}$. }
    \label{fig:additional_settings_B}
\end{figure}

\subsection*{Supplementary Note 4: Charge distributation tensile strain}

In Supplementary \cref{fig:charge_redistribution_sup} we show the average charge redistribution in each direction in the tensile structure (7.8\%). 
We see very similar behavior for the charge along c-direction (Supplementary \cref{fig:charge_redistribution_sup}a)) and perpendicular direction (Supplementary \cref{fig:charge_redistribution_sup}b)) like in the compressive strain (see main text).
As in the compressive case the charge redistribution in the $\beta$-RuCl$_3$ atomic wire, shown by the green curve, sharply decays within few Angstrom in the perpendicular direction.
However, we note a contrast along the chain, where the charge does not order equivalently but has a wave-like behavior (Supplementary \cref{fig:charge_redistribution_sup}c)). 
The spatially resolved charge transfer increases in value close to the Ru sites with less magnetic moment and decreases close to sites with higher magnetic moment.

\begin{figure}[h]
    \centering
    \includegraphics[width=\linewidth]{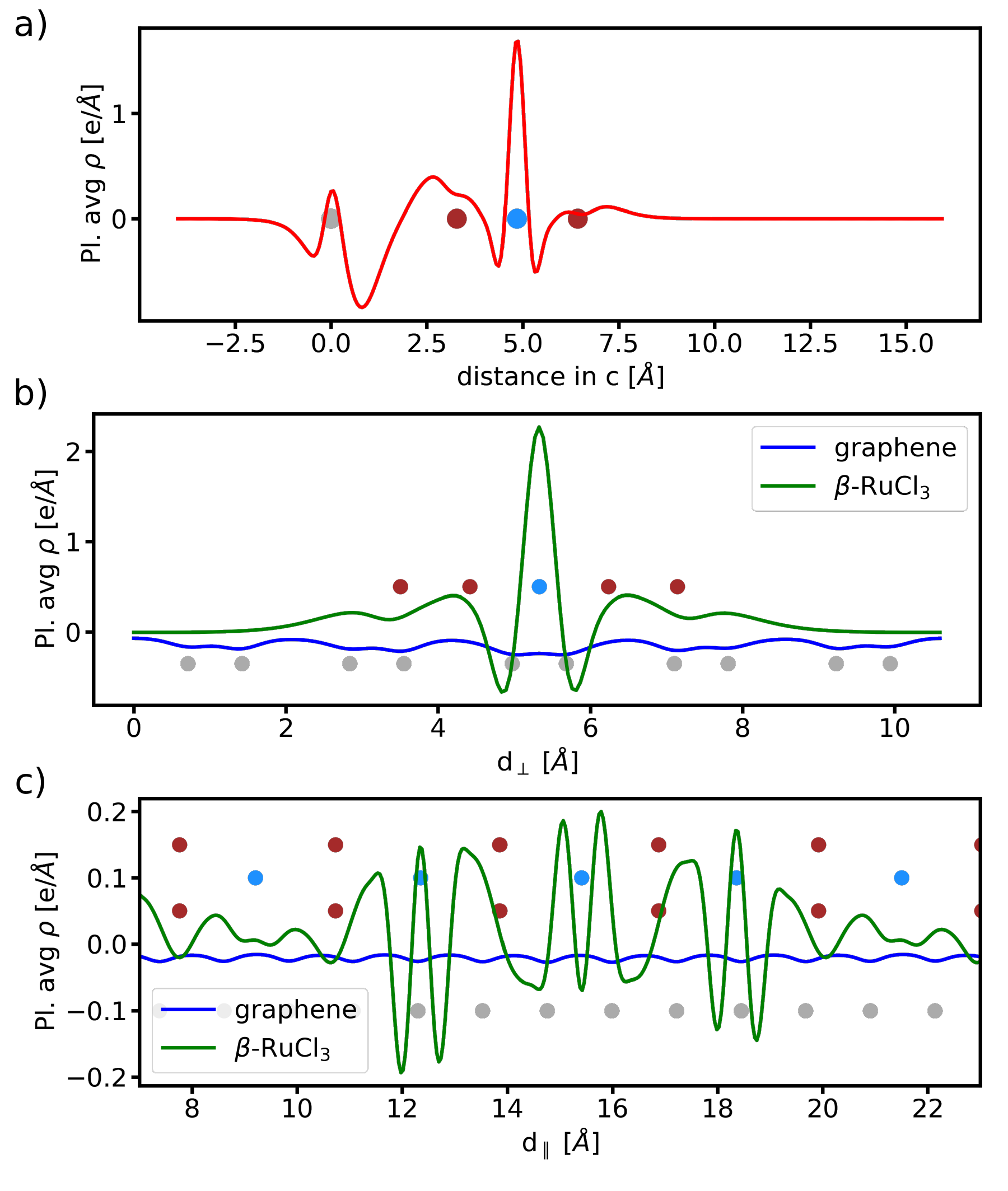}
    \caption{Spatial charge redistribution for the tensile strained heterostructure in each direction. Note: the planer average values $\rho$ are scaled by a factor of 1000.}
    \label{fig:charge_redistribution_sup}
\end{figure}

\subsection*{Supplementary Note 5: Electronic properties, extended energy range}

\begin{figure}
    \centering
    \includegraphics[width=\linewidth]{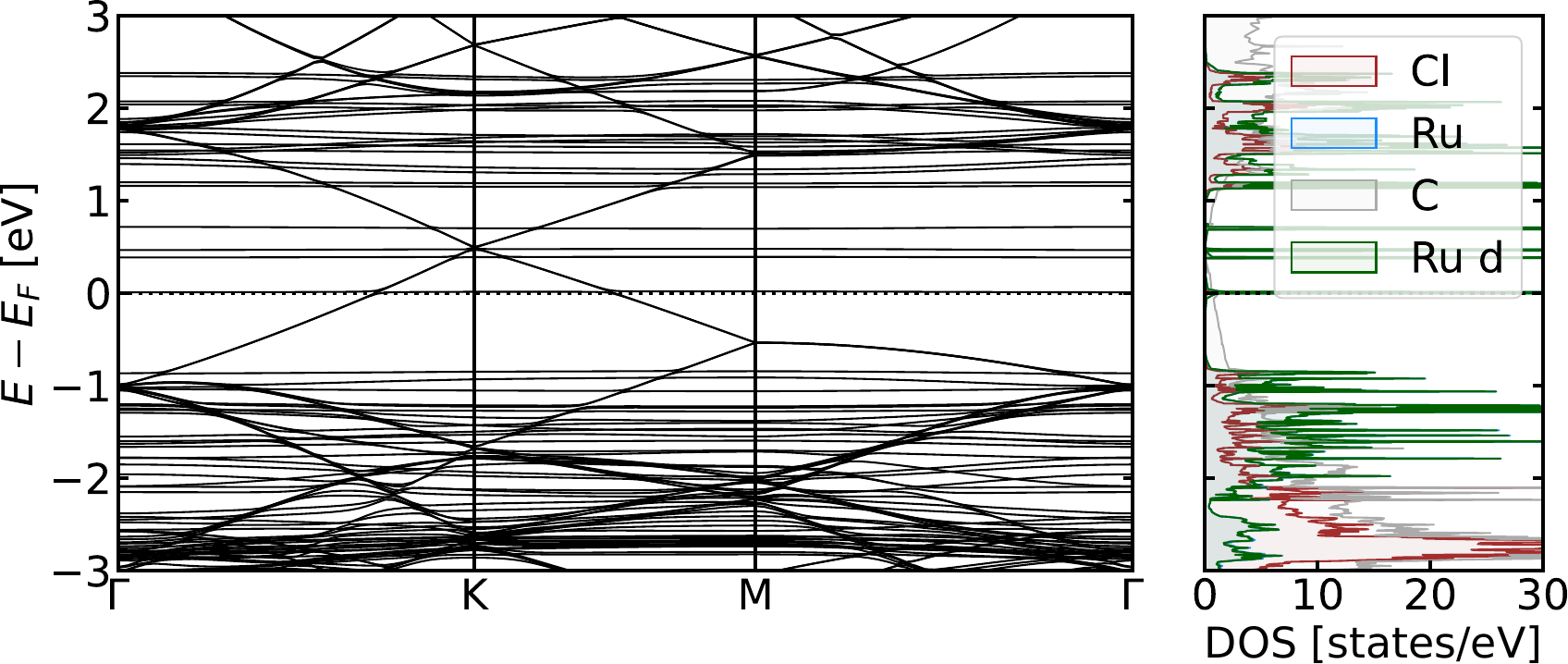}
    \caption{Electronic properties of the relaxed single-chain under tensile strain estimated by DFT+U, with $U-J= U_{\rm eff}=3.0$ eV, in an extended energy window. }
    \label{fig:tensile_extended}
\end{figure}

\begin{figure}[H]
    \centering
    \includegraphics[width=\linewidth]{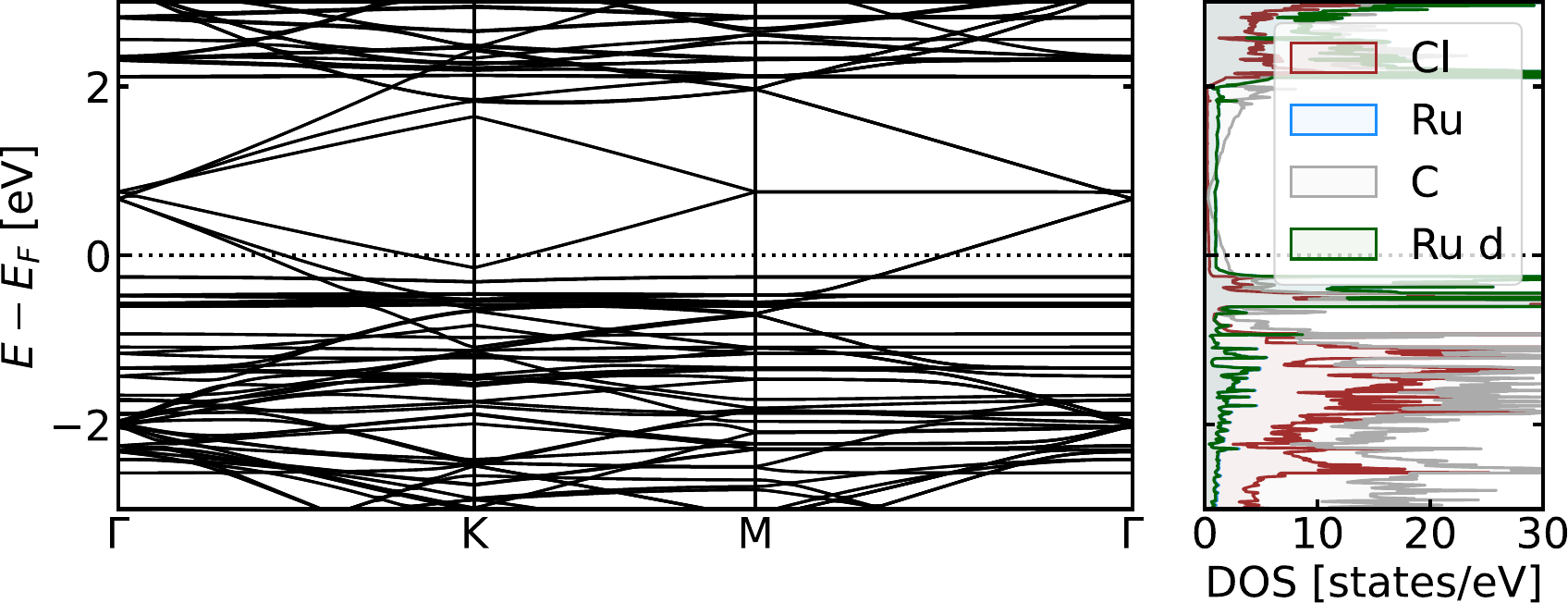}
    \caption{Electronic properties of the relaxed single-chain under compressive strain estimated by DFT+U, with $U-J= U_{\rm eff}=3.0$ eV, in an extended energy window. We see bonding and anti bonding states for the Ru 4d states, at -0.5 eV and 2.5 eV, respectively. }
    \label{fig:compressive_extended}
\end{figure}

In Supplementary~\cref{fig:tensile_extended} and Supplementary~\cref{fig:compressive_extended} we show the electronic band structure and atom and orbital resolved density of states (DOS) within increased energy window for the tensile (7.8\%) and compressive (-13\%) strained structure, respectively.
For compressive strained structure within the increased energy window we see the bonding and antibonding formation around -0.5 eV and 2.5 eV due to the Ru-Ru bond shortening. 

\begin{thebibliography}{50}%
\makeatletter
\providecommand \@ifxundefined [1]{%
 \@ifx{#1\undefined}
}%
\providecommand \@ifnum [1]{%
 \ifnum #1\expandafter \@firstoftwo
 \else \expandafter \@secondoftwo
 \fi
}%
\providecommand \@ifx [1]{%
 \ifx #1\expandafter \@firstoftwo
 \else \expandafter \@secondoftwo
 \fi
}%
\providecommand \natexlab [1]{#1}%
\providecommand \enquote  [1]{``#1''}%
\providecommand \bibnamefont  [1]{#1}%
\providecommand \bibfnamefont [1]{#1}%
\providecommand \citenamefont [1]{#1}%
\providecommand \href@noop [0]{\@secondoftwo}%
\providecommand \href [0]{\begingroup \@sanitize@url \@href}%
\providecommand \@href[1]{\@@startlink{#1}\@@href}%
\providecommand \@@href[1]{\endgroup#1\@@endlink}%
\providecommand \@sanitize@url [0]{\catcode `\\12\catcode `\$12\catcode
  `\&12\catcode `\#12\catcode `\^12\catcode `\_12\catcode `\%12\relax}%
\providecommand \@@startlink[1]{}%
\providecommand \@@endlink[0]{}%
\providecommand \url  [0]{\begingroup\@sanitize@url \@url }%
\providecommand \@url [1]{\endgroup\@href {#1}{\urlprefix }}%
\providecommand \urlprefix  [0]{URL }%
\providecommand \Eprint [0]{\href }%
\providecommand \doibase [0]{https://doi.org/}%
\providecommand \selectlanguage [0]{\@gobble}%
\providecommand \bibinfo  [0]{\@secondoftwo}%
\providecommand \bibfield  [0]{\@secondoftwo}%
\providecommand \translation [1]{[#1]}%
\providecommand \BibitemOpen [0]{}%
\providecommand \bibitemStop [0]{}%
\providecommand \bibitemNoStop [0]{.\EOS\space}%
\providecommand \EOS [0]{\spacefactor3000\relax}%
\providecommand \BibitemShut  [1]{\csname bibitem#1\endcsname}%
\let\auto@bib@innerbib\@empty
\bibitem [{\citenamefont {Geim}\ and\ \citenamefont
  {Grigorieva}(2013)}]{Geim2013}%
  \BibitemOpen
  \bibfield  {author} {\bibinfo {author} {\bibfnamefont {A.~K.}\ \bibnamefont
  {Geim}}\ and\ \bibinfo {author} {\bibfnamefont {I.~V.}\ \bibnamefont
  {Grigorieva}},\ }\bibfield  {title} {\bibinfo {title} {Van der waals
  heterostructures},\ }\href@noop {} {\bibfield  {journal} {\bibinfo  {journal}
  {Nature}\ }\textbf {\bibinfo {volume} {499}},\ \bibinfo {pages} {419}
  (\bibinfo {year} {2013})},\ \bibinfo {note} {perspective}\BibitemShut
  {NoStop}%
\bibitem [{\citenamefont {Novoselov}\ \emph {et~al.}(2016)\citenamefont
  {Novoselov}, \citenamefont {Mishchenko}, \citenamefont {Carvalho},\ and\
  \citenamefont {Castro~Neto}}]{Novoselov2016}%
  \BibitemOpen
  \bibfield  {author} {\bibinfo {author} {\bibfnamefont {K.~S.}\ \bibnamefont
  {Novoselov}}, \bibinfo {author} {\bibfnamefont {A.}~\bibnamefont
  {Mishchenko}}, \bibinfo {author} {\bibfnamefont {A.}~\bibnamefont
  {Carvalho}},\ and\ \bibinfo {author} {\bibfnamefont {A.~H.}\ \bibnamefont
  {Castro~Neto}},\ }\bibfield  {title} {\bibinfo {title} {2d materials and van
  der waals heterostructures},\ }\href@noop {} {\bibfield  {journal} {\bibinfo
  {journal} {Science}\ }\textbf {\bibinfo {volume} {353}} (\bibinfo {year}
  {2016})}\BibitemShut {NoStop}%
\bibitem [{\citenamefont {Yu}\ \emph {et~al.}(2013)\citenamefont {Yu},
  \citenamefont {Li}, \citenamefont {Zhou}, \citenamefont {Chen}, \citenamefont
  {Wang}, \citenamefont {Huang},\ and\ \citenamefont
  {Duan}}]{yu2013vertically}%
  \BibitemOpen
  \bibfield  {author} {\bibinfo {author} {\bibfnamefont {W.~J.}\ \bibnamefont
  {Yu}}, \bibinfo {author} {\bibfnamefont {Z.}~\bibnamefont {Li}}, \bibinfo
  {author} {\bibfnamefont {H.}~\bibnamefont {Zhou}}, \bibinfo {author}
  {\bibfnamefont {Y.}~\bibnamefont {Chen}}, \bibinfo {author} {\bibfnamefont
  {Y.}~\bibnamefont {Wang}}, \bibinfo {author} {\bibfnamefont {Y.}~\bibnamefont
  {Huang}},\ and\ \bibinfo {author} {\bibfnamefont {X.}~\bibnamefont {Duan}},\
  }\bibfield  {title} {\bibinfo {title} {Vertically stacked
  multi-heterostructures of layered materials for logic transistors and
  complementary inverters},\ }\href@noop {} {\bibfield  {journal} {\bibinfo
  {journal} {Nature materials}\ }\textbf {\bibinfo {volume} {12}},\ \bibinfo
  {pages} {246} (\bibinfo {year} {2013})}\BibitemShut {NoStop}%
\bibitem [{\citenamefont {Britnell}\ \emph {et~al.}(2012)\citenamefont
  {Britnell}, \citenamefont {Gorbachev}, \citenamefont {Jalil}, \citenamefont
  {Belle}, \citenamefont {Schedin}, \citenamefont {Mishchenko}, \citenamefont
  {Georgiou}, \citenamefont {Katsnelson}, \citenamefont {Eaves}, \citenamefont
  {Morozov} \emph {et~al.}}]{britnell2012field}%
  \BibitemOpen
  \bibfield  {author} {\bibinfo {author} {\bibfnamefont {L.}~\bibnamefont
  {Britnell}}, \bibinfo {author} {\bibfnamefont {R.}~\bibnamefont {Gorbachev}},
  \bibinfo {author} {\bibfnamefont {R.}~\bibnamefont {Jalil}}, \bibinfo
  {author} {\bibfnamefont {B.}~\bibnamefont {Belle}}, \bibinfo {author}
  {\bibfnamefont {F.}~\bibnamefont {Schedin}}, \bibinfo {author} {\bibfnamefont
  {A.}~\bibnamefont {Mishchenko}}, \bibinfo {author} {\bibfnamefont
  {T.}~\bibnamefont {Georgiou}}, \bibinfo {author} {\bibfnamefont
  {M.}~\bibnamefont {Katsnelson}}, \bibinfo {author} {\bibfnamefont
  {L.}~\bibnamefont {Eaves}}, \bibinfo {author} {\bibfnamefont
  {S.}~\bibnamefont {Morozov}}, \emph {et~al.},\ }\bibfield  {title} {\bibinfo
  {title} {Field-effect tunneling transistor based on vertical graphene
  heterostructures},\ }\href@noop {} {\bibfield  {journal} {\bibinfo  {journal}
  {Science}\ }\textbf {\bibinfo {volume} {335}},\ \bibinfo {pages} {947}
  (\bibinfo {year} {2012})}\BibitemShut {NoStop}%
\bibitem [{\citenamefont {Zhang}\ \emph {et~al.}(2014)\citenamefont {Zhang},
  \citenamefont {Chuu}, \citenamefont {Huang}, \citenamefont {Chen},
  \citenamefont {Tsai}, \citenamefont {Chang}, \citenamefont {Liang},
  \citenamefont {Chen}, \citenamefont {Chueh}, \citenamefont {He} \emph
  {et~al.}}]{zhang2014ultrahigh}%
  \BibitemOpen
  \bibfield  {author} {\bibinfo {author} {\bibfnamefont {W.}~\bibnamefont
  {Zhang}}, \bibinfo {author} {\bibfnamefont {C.-P.}\ \bibnamefont {Chuu}},
  \bibinfo {author} {\bibfnamefont {J.-K.}\ \bibnamefont {Huang}}, \bibinfo
  {author} {\bibfnamefont {C.-H.}\ \bibnamefont {Chen}}, \bibinfo {author}
  {\bibfnamefont {M.-L.}\ \bibnamefont {Tsai}}, \bibinfo {author}
  {\bibfnamefont {Y.-H.}\ \bibnamefont {Chang}}, \bibinfo {author}
  {\bibfnamefont {C.-T.}\ \bibnamefont {Liang}}, \bibinfo {author}
  {\bibfnamefont {Y.-Z.}\ \bibnamefont {Chen}}, \bibinfo {author}
  {\bibfnamefont {Y.-L.}\ \bibnamefont {Chueh}}, \bibinfo {author}
  {\bibfnamefont {J.-H.}\ \bibnamefont {He}}, \emph {et~al.},\ }\bibfield
  {title} {\bibinfo {title} {Ultrahigh-gain photodetectors based on atomically
  thin graphene-mos2 heterostructures},\ }\href@noop {} {\bibfield  {journal}
  {\bibinfo  {journal} {Scientific reports}\ }\textbf {\bibinfo {volume} {4}},\
  \bibinfo {pages} {3826} (\bibinfo {year} {2014})}\BibitemShut {NoStop}%
\bibitem [{\citenamefont {Mehew}\ \emph {et~al.}(2017)\citenamefont {Mehew},
  \citenamefont {Unal}, \citenamefont {Torres~Alonso}, \citenamefont {Jones},
  \citenamefont {Fadhil~Ramadhan}, \citenamefont {Craciun},\ and\ \citenamefont
  {Russo}}]{mehew2017fast}%
  \BibitemOpen
  \bibfield  {author} {\bibinfo {author} {\bibfnamefont {J.~D.}\ \bibnamefont
  {Mehew}}, \bibinfo {author} {\bibfnamefont {S.}~\bibnamefont {Unal}},
  \bibinfo {author} {\bibfnamefont {E.}~\bibnamefont {Torres~Alonso}}, \bibinfo
  {author} {\bibfnamefont {G.~F.}\ \bibnamefont {Jones}}, \bibinfo {author}
  {\bibfnamefont {S.}~\bibnamefont {Fadhil~Ramadhan}}, \bibinfo {author}
  {\bibfnamefont {M.~F.}\ \bibnamefont {Craciun}},\ and\ \bibinfo {author}
  {\bibfnamefont {S.}~\bibnamefont {Russo}},\ }\bibfield  {title} {\bibinfo
  {title} {Fast and highly sensitive ionic-polymer-gated ws2--graphene
  photodetectors},\ }\href@noop {} {\bibfield  {journal} {\bibinfo  {journal}
  {Advanced Materials}\ }\textbf {\bibinfo {volume} {29}},\ \bibinfo {pages}
  {1700222} (\bibinfo {year} {2017})}\BibitemShut {NoStop}%
\bibitem [{\citenamefont {Saraf}\ \emph {et~al.}(2018)\citenamefont {Saraf},
  \citenamefont {Natarajan},\ and\ \citenamefont {Mobin}}]{saraf2018emerging}%
  \BibitemOpen
  \bibfield  {author} {\bibinfo {author} {\bibfnamefont {M.}~\bibnamefont
  {Saraf}}, \bibinfo {author} {\bibfnamefont {K.}~\bibnamefont {Natarajan}},\
  and\ \bibinfo {author} {\bibfnamefont {S.~M.}\ \bibnamefont {Mobin}},\
  }\bibfield  {title} {\bibinfo {title} {Emerging robust heterostructure of
  mos2--rgo for high-performance supercapacitors},\ }\href@noop {} {\bibfield
  {journal} {\bibinfo  {journal} {ACS applied materials \& interfaces}\
  }\textbf {\bibinfo {volume} {10}},\ \bibinfo {pages} {16588} (\bibinfo {year}
  {2018})}\BibitemShut {NoStop}%
\bibitem [{\citenamefont {Ratha}\ and\ \citenamefont
  {Rout}(2013)}]{ratha2013supercapacitor}%
  \BibitemOpen
  \bibfield  {author} {\bibinfo {author} {\bibfnamefont {S.}~\bibnamefont
  {Ratha}}\ and\ \bibinfo {author} {\bibfnamefont {C.~S.}\ \bibnamefont
  {Rout}},\ }\bibfield  {title} {\bibinfo {title} {Supercapacitor electrodes
  based on layered tungsten disulfide-reduced graphene oxide hybrids
  synthesized by a facile hydrothermal method},\ }\href@noop {} {\bibfield
  {journal} {\bibinfo  {journal} {ACS applied materials \& interfaces}\
  }\textbf {\bibinfo {volume} {5}},\ \bibinfo {pages} {11427} (\bibinfo {year}
  {2013})}\BibitemShut {NoStop}%
\bibitem [{\citenamefont {Kirubasankar}\ \emph {et~al.}(2019)\citenamefont
  {Kirubasankar}, \citenamefont {Vijayan},\ and\ \citenamefont
  {Angaiah}}]{kirubasankar2019sonochemical}%
  \BibitemOpen
  \bibfield  {author} {\bibinfo {author} {\bibfnamefont {B.}~\bibnamefont
  {Kirubasankar}}, \bibinfo {author} {\bibfnamefont {S.}~\bibnamefont
  {Vijayan}},\ and\ \bibinfo {author} {\bibfnamefont {S.}~\bibnamefont
  {Angaiah}},\ }\bibfield  {title} {\bibinfo {title} {Sonochemical synthesis of
  a 2d--2d mose 2/graphene nanohybrid electrode material for asymmetric
  supercapacitors},\ }\href@noop {} {\bibfield  {journal} {\bibinfo  {journal}
  {Sustainable Energy \& Fuels}\ }\textbf {\bibinfo {volume} {3}},\ \bibinfo
  {pages} {467} (\bibinfo {year} {2019})}\BibitemShut {NoStop}%
\bibitem [{\citenamefont {Sadeghi}\ \emph {et~al.}(2016)\citenamefont
  {Sadeghi}, \citenamefont {Sangtarash},\ and\ \citenamefont
  {Lambert}}]{sadeghi2016cross}%
  \BibitemOpen
  \bibfield  {author} {\bibinfo {author} {\bibfnamefont {H.}~\bibnamefont
  {Sadeghi}}, \bibinfo {author} {\bibfnamefont {S.}~\bibnamefont
  {Sangtarash}},\ and\ \bibinfo {author} {\bibfnamefont {C.~J.}\ \bibnamefont
  {Lambert}},\ }\bibfield  {title} {\bibinfo {title} {Cross-plane enhanced
  thermoelectricity and phonon suppression in graphene/mos2 van der waals
  heterostructures},\ }\href@noop {} {\bibfield  {journal} {\bibinfo  {journal}
  {2D Materials}\ }\textbf {\bibinfo {volume} {4}},\ \bibinfo {pages} {015012}
  (\bibinfo {year} {2016})}\BibitemShut {NoStop}%
\bibitem [{\citenamefont {Aji}\ \emph {et~al.}(2017)\citenamefont {Aji},
  \citenamefont {Sol{\'\i}s-Fern{\'a}ndez}, \citenamefont {Ji}, \citenamefont
  {Fukuda},\ and\ \citenamefont {Ago}}]{aji2017high}%
  \BibitemOpen
  \bibfield  {author} {\bibinfo {author} {\bibfnamefont {A.~S.}\ \bibnamefont
  {Aji}}, \bibinfo {author} {\bibfnamefont {P.}~\bibnamefont
  {Sol{\'\i}s-Fern{\'a}ndez}}, \bibinfo {author} {\bibfnamefont {H.~G.}\
  \bibnamefont {Ji}}, \bibinfo {author} {\bibfnamefont {K.}~\bibnamefont
  {Fukuda}},\ and\ \bibinfo {author} {\bibfnamefont {H.}~\bibnamefont {Ago}},\
  }\bibfield  {title} {\bibinfo {title} {High mobility ws2 transistors realized
  by multilayer graphene electrodes and application to high responsivity
  flexible photodetectors},\ }\href@noop {} {\bibfield  {journal} {\bibinfo
  {journal} {Advanced Functional Materials}\ }\textbf {\bibinfo {volume}
  {27}},\ \bibinfo {pages} {1703448} (\bibinfo {year} {2017})}\BibitemShut
  {NoStop}%
\bibitem [{\citenamefont {Zhao}\ \emph {et~al.}(2015)\citenamefont {Zhao},
  \citenamefont {Hou}, \citenamefont {Wu}, \citenamefont {He},\ and\
  \citenamefont {Hao}}]{zhao2015preparation}%
  \BibitemOpen
  \bibfield  {author} {\bibinfo {author} {\bibfnamefont {G.}~\bibnamefont
  {Zhao}}, \bibinfo {author} {\bibfnamefont {J.}~\bibnamefont {Hou}}, \bibinfo
  {author} {\bibfnamefont {Y.}~\bibnamefont {Wu}}, \bibinfo {author}
  {\bibfnamefont {J.}~\bibnamefont {He}},\ and\ \bibinfo {author}
  {\bibfnamefont {X.}~\bibnamefont {Hao}},\ }\bibfield  {title} {\bibinfo
  {title} {Preparation of 2d mos2/graphene heterostructure through a monolayer
  intercalation method and its application as an optical modulator in pulsed
  laser generation},\ }\href@noop {} {\bibfield  {journal} {\bibinfo  {journal}
  {Advanced Optical Materials}\ }\textbf {\bibinfo {volume} {3}},\ \bibinfo
  {pages} {937} (\bibinfo {year} {2015})}\BibitemShut {NoStop}%
\bibitem [{\citenamefont {Tseng}\ \emph {et~al.}(2022)\citenamefont {Tseng},
  \citenamefont {Song}, \citenamefont {Jiang}, \citenamefont {Lin},
  \citenamefont {Wang}, \citenamefont {Suh}, \citenamefont {Watanabe},
  \citenamefont {Taniguchi}, \citenamefont {McGuire}, \citenamefont {Xiao}
  \emph {et~al.}}]{tseng2022gate}%
  \BibitemOpen
  \bibfield  {author} {\bibinfo {author} {\bibfnamefont {C.-C.}\ \bibnamefont
  {Tseng}}, \bibinfo {author} {\bibfnamefont {T.}~\bibnamefont {Song}},
  \bibinfo {author} {\bibfnamefont {Q.}~\bibnamefont {Jiang}}, \bibinfo
  {author} {\bibfnamefont {Z.}~\bibnamefont {Lin}}, \bibinfo {author}
  {\bibfnamefont {C.}~\bibnamefont {Wang}}, \bibinfo {author} {\bibfnamefont
  {J.}~\bibnamefont {Suh}}, \bibinfo {author} {\bibfnamefont {K.}~\bibnamefont
  {Watanabe}}, \bibinfo {author} {\bibfnamefont {T.}~\bibnamefont {Taniguchi}},
  \bibinfo {author} {\bibfnamefont {M.~A.}\ \bibnamefont {McGuire}}, \bibinfo
  {author} {\bibfnamefont {D.}~\bibnamefont {Xiao}}, \emph {et~al.},\
  }\bibfield  {title} {\bibinfo {title} {Gate-tunable proximity effects in
  graphene on layered magnetic insulators},\ }\href@noop {} {\bibfield
  {journal} {\bibinfo  {journal} {Nano Letters}\ }\textbf {\bibinfo {volume}
  {22}},\ \bibinfo {pages} {8495} (\bibinfo {year} {2022})}\BibitemShut
  {NoStop}%
\bibitem [{\citenamefont {Zhou}\ \emph {et~al.}(2019)\citenamefont {Zhou},
  \citenamefont {Balgley}, \citenamefont {Lampen-Kelley}, \citenamefont {Yan},
  \citenamefont {Mandrus},\ and\ \citenamefont {Henriksen}}]{zhou2019evidence}%
  \BibitemOpen
  \bibfield  {author} {\bibinfo {author} {\bibfnamefont {B.}~\bibnamefont
  {Zhou}}, \bibinfo {author} {\bibfnamefont {J.}~\bibnamefont {Balgley}},
  \bibinfo {author} {\bibfnamefont {P.}~\bibnamefont {Lampen-Kelley}}, \bibinfo
  {author} {\bibfnamefont {J.-Q.}\ \bibnamefont {Yan}}, \bibinfo {author}
  {\bibfnamefont {D.~G.}\ \bibnamefont {Mandrus}},\ and\ \bibinfo {author}
  {\bibfnamefont {E.~A.}\ \bibnamefont {Henriksen}},\ }\bibfield  {title}
  {\bibinfo {title} {Evidence for charge transfer and proximate magnetism in
  graphene--$\ensuremath{\alpha}\text{\ensuremath{-}}{\mathrm{rucl}}_{3}$
  heterostructures},\ }\href@noop {} {\bibfield  {journal} {\bibinfo  {journal}
  {Phys. Rev. B}\ }\textbf {\bibinfo {volume} {100}},\ \bibinfo {pages}
  {165426} (\bibinfo {year} {2019})}\BibitemShut {NoStop}%
\bibitem [{\citenamefont {Mashhadi}\ \emph {et~al.}(2019)\citenamefont
  {Mashhadi}, \citenamefont {Kim}, \citenamefont {Kim}, \citenamefont {Weber},
  \citenamefont {Taniguchi}, \citenamefont {Watanabe}, \citenamefont {Park},
  \citenamefont {Lotsch}, \citenamefont {Smet}, \citenamefont {Burghard},\ and\
  \citenamefont {Kern}}]{mashhadiArXiv19}%
  \BibitemOpen
  \bibfield  {author} {\bibinfo {author} {\bibfnamefont {S.}~\bibnamefont
  {Mashhadi}}, \bibinfo {author} {\bibfnamefont {Y.}~\bibnamefont {Kim}},
  \bibinfo {author} {\bibfnamefont {J.}~\bibnamefont {Kim}}, \bibinfo {author}
  {\bibfnamefont {D.}~\bibnamefont {Weber}}, \bibinfo {author} {\bibfnamefont
  {T.}~\bibnamefont {Taniguchi}}, \bibinfo {author} {\bibfnamefont
  {K.}~\bibnamefont {Watanabe}}, \bibinfo {author} {\bibfnamefont
  {N.}~\bibnamefont {Park}}, \bibinfo {author} {\bibfnamefont {B.}~\bibnamefont
  {Lotsch}}, \bibinfo {author} {\bibfnamefont {J.~H.}\ \bibnamefont {Smet}},
  \bibinfo {author} {\bibfnamefont {M.}~\bibnamefont {Burghard}},\ and\
  \bibinfo {author} {\bibfnamefont {K.}~\bibnamefont {Kern}},\ }\bibfield
  {title} {\bibinfo {title} {Spin-split band hybridization in graphene
  proximitized with $\alpha$-rucl3 nanosheets},\ }\href@noop {} {\bibfield
  {journal} {\bibinfo  {journal} {Nano Letters}\ }\textbf {\bibinfo {volume}
  {19}},\ \bibinfo {pages} {4659} (\bibinfo {year} {2019})},\ \bibinfo {note}
  {pMID: 31241971}\BibitemShut {NoStop}%
\bibitem [{\citenamefont {Biswas}\ \emph {et~al.}(2019)\citenamefont {Biswas},
  \citenamefont {Li}, \citenamefont {Winter}, \citenamefont {Knolle},\ and\
  \citenamefont {Valent\'{\i}}}]{Biswas_2019_electronic}%
  \BibitemOpen
  \bibfield  {author} {\bibinfo {author} {\bibfnamefont {S.}~\bibnamefont
  {Biswas}}, \bibinfo {author} {\bibfnamefont {Y.}~\bibnamefont {Li}}, \bibinfo
  {author} {\bibfnamefont {S.~M.}\ \bibnamefont {Winter}}, \bibinfo {author}
  {\bibfnamefont {J.}~\bibnamefont {Knolle}},\ and\ \bibinfo {author}
  {\bibfnamefont {R.}~\bibnamefont {Valent\'{\i}}},\ }\bibfield  {title}
  {\bibinfo {title} {Electronic properties of
  $\ensuremath{\alpha}\text{\ensuremath{-}}{\mathrm{rucl}}_{3}$ in proximity to
  graphene},\ }\href@noop {} {\bibfield  {journal} {\bibinfo  {journal} {Phys.
  Rev. Lett.}\ }\textbf {\bibinfo {volume} {123}},\ \bibinfo {pages} {237201}
  (\bibinfo {year} {2019})}\BibitemShut {NoStop}%
\bibitem [{\citenamefont {Rizzo}\ \emph {et~al.}(2020)\citenamefont {Rizzo},
  \citenamefont {Jessen}, \citenamefont {Sun}, \citenamefont {Ruta},
  \citenamefont {Zhang}, \citenamefont {Yan}, \citenamefont {Xian},
  \citenamefont {McLeod}, \citenamefont {Berkowitz}, \citenamefont {Watanabe}
  \emph {et~al.}}]{rizzo2020}%
  \BibitemOpen
  \bibfield  {author} {\bibinfo {author} {\bibfnamefont {D.~J.}\ \bibnamefont
  {Rizzo}}, \bibinfo {author} {\bibfnamefont {B.~S.}\ \bibnamefont {Jessen}},
  \bibinfo {author} {\bibfnamefont {Z.}~\bibnamefont {Sun}}, \bibinfo {author}
  {\bibfnamefont {F.~L.}\ \bibnamefont {Ruta}}, \bibinfo {author}
  {\bibfnamefont {J.}~\bibnamefont {Zhang}}, \bibinfo {author} {\bibfnamefont
  {J.-Q.}\ \bibnamefont {Yan}}, \bibinfo {author} {\bibfnamefont
  {L.}~\bibnamefont {Xian}}, \bibinfo {author} {\bibfnamefont {A.~S.}\
  \bibnamefont {McLeod}}, \bibinfo {author} {\bibfnamefont {M.~E.}\
  \bibnamefont {Berkowitz}}, \bibinfo {author} {\bibfnamefont {K.}~\bibnamefont
  {Watanabe}}, \emph {et~al.},\ }\bibfield  {title} {\bibinfo {title}
  {Charge-transfer plasmon polaritons at graphene/$\alpha$-rucl3 interfaces},\
  }\href@noop {} {\bibfield  {journal} {\bibinfo  {journal} {Nano letters}\
  }\textbf {\bibinfo {volume} {20}},\ \bibinfo {pages} {8438} (\bibinfo {year}
  {2020})}\BibitemShut {NoStop}%
\bibitem [{\citenamefont {Gerber}\ \emph {et~al.}(2020)\citenamefont {Gerber},
  \citenamefont {Yao}, \citenamefont {Arias},\ and\ \citenamefont
  {Kim}}]{gerber2020abinitio}%
  \BibitemOpen
  \bibfield  {author} {\bibinfo {author} {\bibfnamefont {E.}~\bibnamefont
  {Gerber}}, \bibinfo {author} {\bibfnamefont {Y.}~\bibnamefont {Yao}},
  \bibinfo {author} {\bibfnamefont {T.~A.}\ \bibnamefont {Arias}},\ and\
  \bibinfo {author} {\bibfnamefont {E.-A.}\ \bibnamefont {Kim}},\ }\bibfield
  {title} {\bibinfo {title} {Ab initio mismatched interface theory of graphene
  on $\ensuremath{\alpha}\text{\ensuremath{-}}{\mathrm{rucl}}_{3}$: Doping and
  magnetism},\ }\href@noop {} {\bibfield  {journal} {\bibinfo  {journal} {Phys.
  Rev. Lett.}\ }\textbf {\bibinfo {volume} {124}},\ \bibinfo {pages} {106804}
  (\bibinfo {year} {2020})}\BibitemShut {NoStop}%
\bibitem [{\citenamefont {Wang}\ \emph {et~al.}(2020)\citenamefont {Wang},
  \citenamefont {Balgley}, \citenamefont {Gerber}, \citenamefont {Gray},
  \citenamefont {Kumar}, \citenamefont {Lu}, \citenamefont {Yan}, \citenamefont
  {Fereidouni}, \citenamefont {Basnet}, \citenamefont {Yun} \emph
  {et~al.}}]{wang2020}%
  \BibitemOpen
  \bibfield  {author} {\bibinfo {author} {\bibfnamefont {Y.}~\bibnamefont
  {Wang}}, \bibinfo {author} {\bibfnamefont {J.}~\bibnamefont {Balgley}},
  \bibinfo {author} {\bibfnamefont {E.}~\bibnamefont {Gerber}}, \bibinfo
  {author} {\bibfnamefont {M.}~\bibnamefont {Gray}}, \bibinfo {author}
  {\bibfnamefont {N.}~\bibnamefont {Kumar}}, \bibinfo {author} {\bibfnamefont
  {X.}~\bibnamefont {Lu}}, \bibinfo {author} {\bibfnamefont {J.-Q.}\
  \bibnamefont {Yan}}, \bibinfo {author} {\bibfnamefont {A.}~\bibnamefont
  {Fereidouni}}, \bibinfo {author} {\bibfnamefont {R.}~\bibnamefont {Basnet}},
  \bibinfo {author} {\bibfnamefont {S.~J.}\ \bibnamefont {Yun}}, \emph
  {et~al.},\ }\bibfield  {title} {\bibinfo {title} {Modulation doping via a
  two-dimensional atomic crystalline acceptor},\ }\href@noop {} {\bibfield
  {journal} {\bibinfo  {journal} {Nano letters}\ }\textbf {\bibinfo {volume}
  {20}},\ \bibinfo {pages} {8446} (\bibinfo {year} {2020})}\BibitemShut
  {NoStop}%
\bibitem [{\citenamefont {Balgley}\ \emph {et~al.}(2022)\citenamefont
  {Balgley}, \citenamefont {Butler}, \citenamefont {Biswas}, \citenamefont
  {Ge}, \citenamefont {Lagasse}, \citenamefont {Taniguchi}, \citenamefont
  {Watanabe}, \citenamefont {Cothrine}, \citenamefont {Mandrus}, \citenamefont
  {Velasco~Jr} \emph {et~al.}}]{balgley2022}%
  \BibitemOpen
  \bibfield  {author} {\bibinfo {author} {\bibfnamefont {J.}~\bibnamefont
  {Balgley}}, \bibinfo {author} {\bibfnamefont {J.}~\bibnamefont {Butler}},
  \bibinfo {author} {\bibfnamefont {S.}~\bibnamefont {Biswas}}, \bibinfo
  {author} {\bibfnamefont {Z.}~\bibnamefont {Ge}}, \bibinfo {author}
  {\bibfnamefont {S.}~\bibnamefont {Lagasse}}, \bibinfo {author} {\bibfnamefont
  {T.}~\bibnamefont {Taniguchi}}, \bibinfo {author} {\bibfnamefont
  {K.}~\bibnamefont {Watanabe}}, \bibinfo {author} {\bibfnamefont
  {M.}~\bibnamefont {Cothrine}}, \bibinfo {author} {\bibfnamefont {D.~G.}\
  \bibnamefont {Mandrus}}, \bibinfo {author} {\bibfnamefont {J.}~\bibnamefont
  {Velasco~Jr}}, \emph {et~al.},\ }\bibfield  {title} {\bibinfo {title}
  {Ultrasharp lateral p--n junctions in modulation-doped graphene},\
  }\href@noop {} {\bibfield  {journal} {\bibinfo  {journal} {Nano Letters}\
  }\textbf {\bibinfo {volume} {22}},\ \bibinfo {pages} {4124} (\bibinfo {year}
  {2022})}\BibitemShut {NoStop}%
\bibitem [{\citenamefont {Yang}\ \emph {et~al.}(2022)\citenamefont {Yang},
  \citenamefont {Goh}, \citenamefont {Sung}, \citenamefont {Ye}, \citenamefont
  {Biswas}, \citenamefont {Kaib}, \citenamefont {Dhakal}, \citenamefont {Yan},
  \citenamefont {Li}, \citenamefont {Jiang}, \citenamefont {Chen},
  \citenamefont {Lei}, \citenamefont {He}, \citenamefont {Valentí},
  \citenamefont {Winter}, \citenamefont {Hovden},\ and\ \citenamefont
  {Tsen}}]{Yang2022_magnetic}%
  \BibitemOpen
  \bibfield  {author} {\bibinfo {author} {\bibfnamefont {B.}~\bibnamefont
  {Yang}}, \bibinfo {author} {\bibfnamefont {Y.~M.}\ \bibnamefont {Goh}},
  \bibinfo {author} {\bibfnamefont {S.~H.}\ \bibnamefont {Sung}}, \bibinfo
  {author} {\bibfnamefont {G.}~\bibnamefont {Ye}}, \bibinfo {author}
  {\bibfnamefont {S.}~\bibnamefont {Biswas}}, \bibinfo {author} {\bibfnamefont
  {D.~A.~S.}\ \bibnamefont {Kaib}}, \bibinfo {author} {\bibfnamefont
  {R.}~\bibnamefont {Dhakal}}, \bibinfo {author} {\bibfnamefont
  {S.}~\bibnamefont {Yan}}, \bibinfo {author} {\bibfnamefont {C.}~\bibnamefont
  {Li}}, \bibinfo {author} {\bibfnamefont {S.}~\bibnamefont {Jiang}}, \bibinfo
  {author} {\bibfnamefont {F.}~\bibnamefont {Chen}}, \bibinfo {author}
  {\bibfnamefont {H.}~\bibnamefont {Lei}}, \bibinfo {author} {\bibfnamefont
  {R.}~\bibnamefont {He}}, \bibinfo {author} {\bibfnamefont {R.}~\bibnamefont
  {Valentí}}, \bibinfo {author} {\bibfnamefont {S.~M.}\ \bibnamefont
  {Winter}}, \bibinfo {author} {\bibfnamefont {R.}~\bibnamefont {Hovden}},\
  and\ \bibinfo {author} {\bibfnamefont {A.~W.}\ \bibnamefont {Tsen}},\
  }\bibfield  {title} {\bibinfo {title} {Magnetic anisotropy reversal driven by
  structural symmetry-breaking in monolayer $\alpha$-rucl3},\ }\href@noop {}
  {\bibfield  {journal} {\bibinfo  {journal} {Nature Materials}\ }\textbf
  {\bibinfo {volume} {22}},\ \bibinfo {pages} {50–57} (\bibinfo {year}
  {2022})}\BibitemShut {NoStop}%
\bibitem [{\citenamefont {Rossi}\ \emph {et~al.}(2023)\citenamefont {Rossi},
  \citenamefont {Johnson}, \citenamefont {Balgley}, \citenamefont {Thomas},
  \citenamefont {Francaviglia}, \citenamefont {Dettori}, \citenamefont
  {Schmid}, \citenamefont {Watanabe}, \citenamefont {Taniguchi}, \citenamefont
  {Cothrine} \emph {et~al.}}]{rossi2023direct}%
  \BibitemOpen
  \bibfield  {author} {\bibinfo {author} {\bibfnamefont {A.}~\bibnamefont
  {Rossi}}, \bibinfo {author} {\bibfnamefont {C.}~\bibnamefont {Johnson}},
  \bibinfo {author} {\bibfnamefont {J.}~\bibnamefont {Balgley}}, \bibinfo
  {author} {\bibfnamefont {J.~C.}\ \bibnamefont {Thomas}}, \bibinfo {author}
  {\bibfnamefont {L.}~\bibnamefont {Francaviglia}}, \bibinfo {author}
  {\bibfnamefont {R.}~\bibnamefont {Dettori}}, \bibinfo {author} {\bibfnamefont
  {A.~K.}\ \bibnamefont {Schmid}}, \bibinfo {author} {\bibfnamefont
  {K.}~\bibnamefont {Watanabe}}, \bibinfo {author} {\bibfnamefont
  {T.}~\bibnamefont {Taniguchi}}, \bibinfo {author} {\bibfnamefont
  {M.}~\bibnamefont {Cothrine}}, \emph {et~al.},\ }\bibfield  {title} {\bibinfo
  {title} {Direct visualization of the charge transfer in a
  graphene/$\alpha$-rucl3 heterostructure via angle-resolved photoemission
  spectroscopy},\ }\href@noop {} {\bibfield  {journal} {\bibinfo  {journal}
  {Nano Letters}\ }\textbf {\bibinfo {volume} {23}},\ \bibinfo {pages} {8000}
  (\bibinfo {year} {2023})}\BibitemShut {NoStop}%
\bibitem [{\citenamefont {Leeb}\ \emph {et~al.}(2021)\citenamefont {Leeb},
  \citenamefont {Polyudov}, \citenamefont {Mashhadi}, \citenamefont {Biswas},
  \citenamefont {Valent{\'\i}}, \citenamefont {Burghard},\ and\ \citenamefont
  {Knolle}}]{leeb2021}%
  \BibitemOpen
  \bibfield  {author} {\bibinfo {author} {\bibfnamefont {V.}~\bibnamefont
  {Leeb}}, \bibinfo {author} {\bibfnamefont {K.}~\bibnamefont {Polyudov}},
  \bibinfo {author} {\bibfnamefont {S.}~\bibnamefont {Mashhadi}}, \bibinfo
  {author} {\bibfnamefont {S.}~\bibnamefont {Biswas}}, \bibinfo {author}
  {\bibfnamefont {R.}~\bibnamefont {Valent{\'\i}}}, \bibinfo {author}
  {\bibfnamefont {M.}~\bibnamefont {Burghard}},\ and\ \bibinfo {author}
  {\bibfnamefont {J.}~\bibnamefont {Knolle}},\ }\bibfield  {title} {\bibinfo
  {title} {Anomalous quantum oscillations in a heterostructure of graphene on a
  proximate quantum spin liquid},\ }\href@noop {} {\bibfield  {journal}
  {\bibinfo  {journal} {Physical Review Letters}\ }\textbf {\bibinfo {volume}
  {126}},\ \bibinfo {pages} {097201} (\bibinfo {year} {2021})}\BibitemShut
  {NoStop}%
\bibitem [{\citenamefont {Asaba}\ \emph {et~al.}(2023)\citenamefont {Asaba},
  \citenamefont {Peng}, \citenamefont {Ono}, \citenamefont {Akutagawa},
  \citenamefont {Tanaka}, \citenamefont {Murayama}, \citenamefont {Suetsugu},
  \citenamefont {Razpopov}, \citenamefont {Kasahara}, \citenamefont
  {Terashima}, \citenamefont {Kohsaka}, \citenamefont {Shibauchi},
  \citenamefont {Ichikawa}, \citenamefont {Valentí}, \citenamefont {ichi
  Sasa},\ and\ \citenamefont {Matsuda}}]{tomoya2023_growth}%
  \BibitemOpen
  \bibfield  {author} {\bibinfo {author} {\bibfnamefont {T.}~\bibnamefont
  {Asaba}}, \bibinfo {author} {\bibfnamefont {L.}~\bibnamefont {Peng}},
  \bibinfo {author} {\bibfnamefont {T.}~\bibnamefont {Ono}}, \bibinfo {author}
  {\bibfnamefont {S.}~\bibnamefont {Akutagawa}}, \bibinfo {author}
  {\bibfnamefont {I.}~\bibnamefont {Tanaka}}, \bibinfo {author} {\bibfnamefont
  {H.}~\bibnamefont {Murayama}}, \bibinfo {author} {\bibfnamefont
  {S.}~\bibnamefont {Suetsugu}}, \bibinfo {author} {\bibfnamefont
  {A.}~\bibnamefont {Razpopov}}, \bibinfo {author} {\bibfnamefont
  {Y.}~\bibnamefont {Kasahara}}, \bibinfo {author} {\bibfnamefont
  {T.}~\bibnamefont {Terashima}}, \bibinfo {author} {\bibfnamefont
  {Y.}~\bibnamefont {Kohsaka}}, \bibinfo {author} {\bibfnamefont
  {T.}~\bibnamefont {Shibauchi}}, \bibinfo {author} {\bibfnamefont
  {M.}~\bibnamefont {Ichikawa}}, \bibinfo {author} {\bibfnamefont
  {R.}~\bibnamefont {Valentí}}, \bibinfo {author} {\bibfnamefont
  {S.}~\bibnamefont {ichi Sasa}},\ and\ \bibinfo {author} {\bibfnamefont
  {Y.}~\bibnamefont {Matsuda}},\ }\bibfield  {title} {\bibinfo {title} {Growth
  of self-integrated atomic quantum wires and junctions of a mott
  semiconductor},\ }\href@noop {} {\bibfield  {journal} {\bibinfo  {journal}
  {Science Advances}\ }\textbf {\bibinfo {volume} {9}},\ \bibinfo {pages}
  {eabq5561} (\bibinfo {year} {2023})}\BibitemShut {NoStop}%
\bibitem [{\citenamefont {Imai}\ \emph {et~al.}(2022)\citenamefont {Imai},
  \citenamefont {Nawa}, \citenamefont {Shimizu}, \citenamefont {Yamada},
  \citenamefont {Fujihara}, \citenamefont {Aoyama}, \citenamefont {Takahashi},
  \citenamefont {Okuyama}, \citenamefont {Ohashi}, \citenamefont {Hagihala}
  \emph {et~al.}}]{imai2022zigzag}%
  \BibitemOpen
  \bibfield  {author} {\bibinfo {author} {\bibfnamefont {Y.}~\bibnamefont
  {Imai}}, \bibinfo {author} {\bibfnamefont {K.}~\bibnamefont {Nawa}}, \bibinfo
  {author} {\bibfnamefont {Y.}~\bibnamefont {Shimizu}}, \bibinfo {author}
  {\bibfnamefont {W.}~\bibnamefont {Yamada}}, \bibinfo {author} {\bibfnamefont
  {H.}~\bibnamefont {Fujihara}}, \bibinfo {author} {\bibfnamefont
  {T.}~\bibnamefont {Aoyama}}, \bibinfo {author} {\bibfnamefont
  {R.}~\bibnamefont {Takahashi}}, \bibinfo {author} {\bibfnamefont
  {D.}~\bibnamefont {Okuyama}}, \bibinfo {author} {\bibfnamefont
  {T.}~\bibnamefont {Ohashi}}, \bibinfo {author} {\bibfnamefont
  {M.}~\bibnamefont {Hagihala}}, \emph {et~al.},\ }\bibfield  {title} {\bibinfo
  {title} {Zigzag magnetic order in the kitaev spin-liquid candidate material
  rubr 3 with a honeycomb lattice},\ }\href@noop {} {\bibfield  {journal}
  {\bibinfo  {journal} {Physical Review B}\ }\textbf {\bibinfo {volume}
  {105}},\ \bibinfo {pages} {L041112} (\bibinfo {year} {2022})}\BibitemShut
  {NoStop}%
\bibitem [{\citenamefont {Nawa}\ \emph {et~al.}(2021)\citenamefont {Nawa},
  \citenamefont {Imai}, \citenamefont {Yamaji}, \citenamefont {Fujihara},
  \citenamefont {Yamada}, \citenamefont {Takahashi}, \citenamefont {Hiraoka},
  \citenamefont {Hagihala}, \citenamefont {Torii}, \citenamefont {Aoyama} \emph
  {et~al.}}]{nawa2021strongly}%
  \BibitemOpen
  \bibfield  {author} {\bibinfo {author} {\bibfnamefont {K.}~\bibnamefont
  {Nawa}}, \bibinfo {author} {\bibfnamefont {Y.}~\bibnamefont {Imai}}, \bibinfo
  {author} {\bibfnamefont {Y.}~\bibnamefont {Yamaji}}, \bibinfo {author}
  {\bibfnamefont {H.}~\bibnamefont {Fujihara}}, \bibinfo {author}
  {\bibfnamefont {W.}~\bibnamefont {Yamada}}, \bibinfo {author} {\bibfnamefont
  {R.}~\bibnamefont {Takahashi}}, \bibinfo {author} {\bibfnamefont
  {T.}~\bibnamefont {Hiraoka}}, \bibinfo {author} {\bibfnamefont
  {M.}~\bibnamefont {Hagihala}}, \bibinfo {author} {\bibfnamefont
  {S.}~\bibnamefont {Torii}}, \bibinfo {author} {\bibfnamefont
  {T.}~\bibnamefont {Aoyama}}, \emph {et~al.},\ }\bibfield  {title} {\bibinfo
  {title} {Strongly electron-correlated semimetal rui3 with a layered honeycomb
  structure},\ }\href@noop {} {\bibfield  {journal} {\bibinfo  {journal}
  {Journal of the Physical Society of Japan}\ }\textbf {\bibinfo {volume}
  {90}},\ \bibinfo {pages} {123703} (\bibinfo {year} {2021})}\BibitemShut
  {NoStop}%
\bibitem [{\citenamefont {Ni}\ \emph {et~al.}(2022)\citenamefont {Ni},
  \citenamefont {Gui}, \citenamefont {Powderly},\ and\ \citenamefont
  {Cava}}]{ni2022honeycomb}%
  \BibitemOpen
  \bibfield  {author} {\bibinfo {author} {\bibfnamefont {D.}~\bibnamefont
  {Ni}}, \bibinfo {author} {\bibfnamefont {X.}~\bibnamefont {Gui}}, \bibinfo
  {author} {\bibfnamefont {K.~M.}\ \bibnamefont {Powderly}},\ and\ \bibinfo
  {author} {\bibfnamefont {R.~J.}\ \bibnamefont {Cava}},\ }\bibfield  {title}
  {\bibinfo {title} {Honeycomb-structure rui3, a new quantum material related
  to $\alpha$-rucl3},\ }\href@noop {} {\bibfield  {journal} {\bibinfo
  {journal} {Advanced Materials}\ }\textbf {\bibinfo {volume} {34}},\ \bibinfo
  {pages} {2106831} (\bibinfo {year} {2022})}\BibitemShut {NoStop}%
\bibitem [{\citenamefont {Winter}\ \emph {et~al.}(2017)\citenamefont {Winter},
  \citenamefont {Tsirlin}, \citenamefont {Daghofer}, \citenamefont {van~den
  Brink}, \citenamefont {Singh}, \citenamefont {Gegenwart},\ and\ \citenamefont
  {Valenti}}]{winter2017models}%
  \BibitemOpen
  \bibfield  {author} {\bibinfo {author} {\bibfnamefont {S.~M.}\ \bibnamefont
  {Winter}}, \bibinfo {author} {\bibfnamefont {A.~A.}\ \bibnamefont {Tsirlin}},
  \bibinfo {author} {\bibfnamefont {M.}~\bibnamefont {Daghofer}}, \bibinfo
  {author} {\bibfnamefont {J.}~\bibnamefont {van~den Brink}}, \bibinfo {author}
  {\bibfnamefont {Y.}~\bibnamefont {Singh}}, \bibinfo {author} {\bibfnamefont
  {P.}~\bibnamefont {Gegenwart}},\ and\ \bibinfo {author} {\bibfnamefont
  {R.}~\bibnamefont {Valenti}},\ }\bibfield  {title} {\bibinfo {title} {Models
  and materials for generalized kitaev magnetism},\ }\href@noop {} {\bibfield
  {journal} {\bibinfo  {journal} {Journal of Physics: Condensed Matter}\
  }\textbf {\bibinfo {volume} {29}},\ \bibinfo {pages} {493002} (\bibinfo
  {year} {2017})}\BibitemShut {NoStop}%
\bibitem [{\citenamefont {Motome}\ and\ \citenamefont
  {Nasu}(2020)}]{motome2020hunting}%
  \BibitemOpen
  \bibfield  {author} {\bibinfo {author} {\bibfnamefont {Y.}~\bibnamefont
  {Motome}}\ and\ \bibinfo {author} {\bibfnamefont {J.}~\bibnamefont {Nasu}},\
  }\bibfield  {title} {\bibinfo {title} {Hunting majorana fermions in kitaev
  magnets},\ }\href@noop {} {\bibfield  {journal} {\bibinfo  {journal} {Journal
  of the Physical Society of Japan}\ }\textbf {\bibinfo {volume} {89}},\
  \bibinfo {pages} {012002} (\bibinfo {year} {2020})}\BibitemShut {NoStop}%
\bibitem [{\citenamefont {Kaib}\ \emph {et~al.}(2022)\citenamefont {Kaib},
  \citenamefont {Riedl}, \citenamefont {Razpopov}, \citenamefont {Li},
  \citenamefont {Backes}, \citenamefont {Mazin},\ and\ \citenamefont
  {Valent{\'\i}}}]{kaib2022electronic}%
  \BibitemOpen
  \bibfield  {author} {\bibinfo {author} {\bibfnamefont {D.~A.}\ \bibnamefont
  {Kaib}}, \bibinfo {author} {\bibfnamefont {K.}~\bibnamefont {Riedl}},
  \bibinfo {author} {\bibfnamefont {A.}~\bibnamefont {Razpopov}}, \bibinfo
  {author} {\bibfnamefont {Y.}~\bibnamefont {Li}}, \bibinfo {author}
  {\bibfnamefont {S.}~\bibnamefont {Backes}}, \bibinfo {author} {\bibfnamefont
  {I.~I.}\ \bibnamefont {Mazin}},\ and\ \bibinfo {author} {\bibfnamefont
  {R.}~\bibnamefont {Valent{\'\i}}},\ }\bibfield  {title} {\bibinfo {title}
  {Electronic and magnetic properties of the rux3 (x= cl, br, i) family: two
  siblings—and a cousin?},\ }\href@noop {} {\bibfield  {journal} {\bibinfo
  {journal} {npj Quantum Materials}\ }\textbf {\bibinfo {volume} {7}},\
  \bibinfo {pages} {75} (\bibinfo {year} {2022})}\BibitemShut {NoStop}%
\bibitem [{\citenamefont {Kim}\ \emph {et~al.}(2022)\citenamefont {Kim},
  \citenamefont {Yuan},\ and\ \citenamefont {Kim}}]{kim2022alpha}%
  \BibitemOpen
  \bibfield  {author} {\bibinfo {author} {\bibfnamefont {S.}~\bibnamefont
  {Kim}}, \bibinfo {author} {\bibfnamefont {B.}~\bibnamefont {Yuan}},\ and\
  \bibinfo {author} {\bibfnamefont {Y.-J.}\ \bibnamefont {Kim}},\ }\bibfield
  {title} {\bibinfo {title} {$\alpha$-rucl3 and other kitaev materials},\
  }\href@noop {} {\bibfield  {journal} {\bibinfo  {journal} {APL Materials}\
  }\textbf {\bibinfo {volume} {10}} (\bibinfo {year} {2022})}\BibitemShut
  {NoStop}%
\bibitem [{\citenamefont {Trebst}\ and\ \citenamefont
  {Hickey}(2022)}]{trebst2022kitaev}%
  \BibitemOpen
  \bibfield  {author} {\bibinfo {author} {\bibfnamefont {S.}~\bibnamefont
  {Trebst}}\ and\ \bibinfo {author} {\bibfnamefont {C.}~\bibnamefont
  {Hickey}},\ }\bibfield  {title} {\bibinfo {title} {Kitaev materials},\
  }\href@noop {} {\bibfield  {journal} {\bibinfo  {journal} {Physics Reports}\
  }\textbf {\bibinfo {volume} {950}},\ \bibinfo {pages} {1} (\bibinfo {year}
  {2022})}\BibitemShut {NoStop}%
\bibitem [{\citenamefont {Ahn}\ \emph {et~al.}(2024)\citenamefont {Ahn},
  \citenamefont {Guo}, \citenamefont {Son}, \citenamefont {Sun},\ and\
  \citenamefont {Zhao}}]{ahn2024progress}%
  \BibitemOpen
  \bibfield  {author} {\bibinfo {author} {\bibfnamefont {Y.}~\bibnamefont
  {Ahn}}, \bibinfo {author} {\bibfnamefont {X.}~\bibnamefont {Guo}}, \bibinfo
  {author} {\bibfnamefont {S.}~\bibnamefont {Son}}, \bibinfo {author}
  {\bibfnamefont {Z.}~\bibnamefont {Sun}},\ and\ \bibinfo {author}
  {\bibfnamefont {L.}~\bibnamefont {Zhao}},\ }\bibfield  {title} {\bibinfo
  {title} {Progress and prospects in two-dimensional magnetism of van der waals
  materials},\ }\href@noop {} {\bibfield  {journal} {\bibinfo  {journal}
  {Progress in Quantum Electronics}\ ,\ \bibinfo {pages} {100498}} (\bibinfo
  {year} {2024})}\BibitemShut {NoStop}%
\bibitem [{\citenamefont {Bl\"ochl}(1994)}]{Bloechl1994}%
  \BibitemOpen
  \bibfield  {author} {\bibinfo {author} {\bibfnamefont {P.~E.}\ \bibnamefont
  {Bl\"ochl}},\ }\bibfield  {title} {\bibinfo {title} {Projector augmented-wave
  method},\ }\href@noop {} {\bibfield  {journal} {\bibinfo  {journal} {Phys.
  Rev. B}\ }\textbf {\bibinfo {volume} {50}},\ \bibinfo {pages} {17953}
  (\bibinfo {year} {1994})}\BibitemShut {NoStop}%
\bibitem [{\citenamefont {Kresse}\ and\ \citenamefont {Hafner}(1993)}]{Kresse}%
  \BibitemOpen
  \bibfield  {author} {\bibinfo {author} {\bibfnamefont {G.}~\bibnamefont
  {Kresse}}\ and\ \bibinfo {author} {\bibfnamefont {J.}~\bibnamefont
  {Hafner}},\ }\bibfield  {title} {\bibinfo {title} {Ab initio molecular
  dynamics for liquid metals},\ }\href@noop {} {\bibfield  {journal} {\bibinfo
  {journal} {Phys. Rev. B}\ }\textbf {\bibinfo {volume} {47}},\ \bibinfo
  {pages} {558} (\bibinfo {year} {1993})}\BibitemShut {NoStop}%
\bibitem [{\citenamefont {Perdew}\ \emph {et~al.}(1996)\citenamefont {Perdew},
  \citenamefont {Burke},\ and\ \citenamefont {Ernzerhof}}]{Perdew1996}%
  \BibitemOpen
  \bibfield  {author} {\bibinfo {author} {\bibfnamefont {J.~P.}\ \bibnamefont
  {Perdew}}, \bibinfo {author} {\bibfnamefont {K.}~\bibnamefont {Burke}},\ and\
  \bibinfo {author} {\bibfnamefont {M.}~\bibnamefont {Ernzerhof}},\ }\bibfield
  {title} {\bibinfo {title} {Generalized gradient approximation made simple},\
  }\href@noop {} {\bibfield  {journal} {\bibinfo  {journal} {Phys. Rev. Lett.}\
  }\textbf {\bibinfo {volume} {77}},\ \bibinfo {pages} {3865} (\bibinfo {year}
  {1996})}\BibitemShut {NoStop}%
\bibitem [{\citenamefont {Dudarev}\ \emph {et~al.}(1998)\citenamefont
  {Dudarev}, \citenamefont {Botton}, \citenamefont {Savrasov}, \citenamefont
  {Humphreys},\ and\ \citenamefont {Sutton}}]{Dudarev}%
  \BibitemOpen
  \bibfield  {author} {\bibinfo {author} {\bibfnamefont {S.~L.}\ \bibnamefont
  {Dudarev}}, \bibinfo {author} {\bibfnamefont {G.~A.}\ \bibnamefont {Botton}},
  \bibinfo {author} {\bibfnamefont {S.~Y.}\ \bibnamefont {Savrasov}}, \bibinfo
  {author} {\bibfnamefont {C.~J.}\ \bibnamefont {Humphreys}},\ and\ \bibinfo
  {author} {\bibfnamefont {A.~P.}\ \bibnamefont {Sutton}},\ }\bibfield  {title}
  {\bibinfo {title} {Electron-energy-loss spectra and the structural stability
  of nickel oxide: An lsda+u study},\ }\href@noop {} {\bibfield  {journal}
  {\bibinfo  {journal} {Phys. Rev. B}\ }\textbf {\bibinfo {volume} {57}},\
  \bibinfo {pages} {1505} (\bibinfo {year} {1998})}\BibitemShut {NoStop}%
\bibitem [{\citenamefont {Grimme}(2006)}]{Grimme}%
  \BibitemOpen
  \bibfield  {author} {\bibinfo {author} {\bibfnamefont {S.}~\bibnamefont
  {Grimme}},\ }\bibfield  {title} {\bibinfo {title} {Semiempirical gga-type
  density functional constructedwith a long-range dispersion correction},\
  }\href@noop {} {\bibfield  {journal} {\bibinfo  {journal} {Journal of Physics
  and Chemistry of Solids}\ }\textbf {\bibinfo {volume} {27}},\ \bibinfo
  {pages} {1787} (\bibinfo {year} {2006})}\BibitemShut {NoStop}%
\bibitem [{\citenamefont {Fletcher}\ \emph {et~al.}(1967)\citenamefont
  {Fletcher}, \citenamefont {Gardner}, \citenamefont {Fox},\ and\ \citenamefont
  {Topping}}]{x-ray_fletcher_1967}%
  \BibitemOpen
  \bibfield  {author} {\bibinfo {author} {\bibfnamefont {J.~M.}\ \bibnamefont
  {Fletcher}}, \bibinfo {author} {\bibfnamefont {W.~E.}\ \bibnamefont
  {Gardner}}, \bibinfo {author} {\bibfnamefont {A.~C.}\ \bibnamefont {Fox}},\
  and\ \bibinfo {author} {\bibfnamefont {G.}~\bibnamefont {Topping}},\
  }\bibfield  {title} {\bibinfo {title} {X-ray{,} infrared{,} and magnetic
  studies of $\alpha$- and $\beta$-ruthenium trichloride},\ }\href@noop {}
  {\bibfield  {journal} {\bibinfo  {journal} {J. Chem. Soc. A}\ ,\ \bibinfo
  {pages} {1038}} (\bibinfo {year} {1967})}\BibitemShut {NoStop}%
\bibitem [{\citenamefont {Kahn}(2016)}]{kahn2016fermi}%
  \BibitemOpen
  \bibfield  {author} {\bibinfo {author} {\bibfnamefont {A.}~\bibnamefont
  {Kahn}},\ }\bibfield  {title} {\bibinfo {title} {Fermi level, work function
  and vacuum level},\ }\href@noop {} {\bibfield  {journal} {\bibinfo  {journal}
  {Materials Horizons}\ }\textbf {\bibinfo {volume} {3}},\ \bibinfo {pages} {7}
  (\bibinfo {year} {2016})}\BibitemShut {NoStop}%
\bibitem [{\citenamefont {Kim}\ and\ \citenamefont
  {Choi}(2021)}]{Kim2021thickness}%
  \BibitemOpen
  \bibfield  {author} {\bibinfo {author} {\bibfnamefont {H.-g.}\ \bibnamefont
  {Kim}}\ and\ \bibinfo {author} {\bibfnamefont {H.~J.}\ \bibnamefont {Choi}},\
  }\bibfield  {title} {\bibinfo {title} {Thickness dependence of work function,
  ionization energy, and electron affinity of mo and w dichalcogenides from dft
  and gw calculations},\ }\href@noop {} {\bibfield  {journal} {\bibinfo
  {journal} {Phys. Rev. B}\ }\textbf {\bibinfo {volume} {103}},\ \bibinfo
  {pages} {085404} (\bibinfo {year} {2021})}\BibitemShut {NoStop}%
\bibitem [{\citenamefont {Henkelman}\ \emph {et~al.}(2006)\citenamefont
  {Henkelman}, \citenamefont {Arnaldsson},\ and\ \citenamefont
  {Jónsson}}]{HENKELMAN_2006_a_fast}%
  \BibitemOpen
  \bibfield  {author} {\bibinfo {author} {\bibfnamefont {G.}~\bibnamefont
  {Henkelman}}, \bibinfo {author} {\bibfnamefont {A.}~\bibnamefont
  {Arnaldsson}},\ and\ \bibinfo {author} {\bibfnamefont {H.}~\bibnamefont
  {Jónsson}},\ }\bibfield  {title} {\bibinfo {title} {A fast and robust
  algorithm for bader decomposition of charge density},\ }\href@noop {}
  {\bibfield  {journal} {\bibinfo  {journal} {Computational Materials Science}\
  }\textbf {\bibinfo {volume} {36}},\ \bibinfo {pages} {354} (\bibinfo {year}
  {2006})}\BibitemShut {NoStop}%
\bibitem [{\citenamefont {Rut'kov}\ \emph {et~al.}(2020)\citenamefont
  {Rut'kov}, \citenamefont {Afanas'eva},\ and\ \citenamefont
  {Gall}}]{Rutkov_2020_graphene}%
  \BibitemOpen
  \bibfield  {author} {\bibinfo {author} {\bibfnamefont {E.}~\bibnamefont
  {Rut'kov}}, \bibinfo {author} {\bibfnamefont {E.}~\bibnamefont
  {Afanas'eva}},\ and\ \bibinfo {author} {\bibfnamefont {N.}~\bibnamefont
  {Gall}},\ }\bibfield  {title} {\bibinfo {title} {Graphene and graphite work
  function depending on layer number on re},\ }\href@noop {} {\bibfield
  {journal} {\bibinfo  {journal} {Diamond and Related Materials}\ }\textbf
  {\bibinfo {volume} {101}},\ \bibinfo {pages} {107576} (\bibinfo {year}
  {2020})}\BibitemShut {NoStop}%
\bibitem [{\citenamefont {Pollini}(1996)}]{pollini_1996_electronic}%
  \BibitemOpen
  \bibfield  {author} {\bibinfo {author} {\bibfnamefont {I.}~\bibnamefont
  {Pollini}},\ }\bibfield  {title} {\bibinfo {title} {Electronic properties of
  the narrow-band material \ensuremath{\alpha}-${\mathrm{rucl}}_{3}$},\
  }\href@noop {} {\bibfield  {journal} {\bibinfo  {journal} {Phys. Rev. B}\
  }\textbf {\bibinfo {volume} {53}},\ \bibinfo {pages} {12769} (\bibinfo {year}
  {1996})}\BibitemShut {NoStop}%
\bibitem [{\citenamefont {Klaproth}\ \emph {et~al.}(2022)\citenamefont
  {Klaproth}, \citenamefont {Grönke}, \citenamefont {Hampel}, \citenamefont
  {Knupfer}, \citenamefont {Büchner}, \citenamefont {Isaeva}, \citenamefont
  {Doert},\ and\ \citenamefont {Koitzsch}}]{Klaproth_2022_work_fuction}%
  \BibitemOpen
  \bibfield  {author} {\bibinfo {author} {\bibfnamefont {T.}~\bibnamefont
  {Klaproth}}, \bibinfo {author} {\bibfnamefont {M.}~\bibnamefont {Grönke}},
  \bibinfo {author} {\bibfnamefont {S.}~\bibnamefont {Hampel}}, \bibinfo
  {author} {\bibfnamefont {M.}~\bibnamefont {Knupfer}}, \bibinfo {author}
  {\bibfnamefont {B.}~\bibnamefont {Büchner}}, \bibinfo {author}
  {\bibfnamefont {A.}~\bibnamefont {Isaeva}}, \bibinfo {author} {\bibfnamefont
  {T.}~\bibnamefont {Doert}},\ and\ \bibinfo {author} {\bibfnamefont
  {A.}~\bibnamefont {Koitzsch}},\ }\bibfield  {title} {\bibinfo {title} {Work
  function engineering of thin $\alpha$-rucl3 by argon sputtering},\
  }\href@noop {} {\bibfield  {journal} {\bibinfo  {journal} {Advanced Materials
  Interfaces}\ }\textbf {\bibinfo {volume} {9}},\ \bibinfo {pages} {2200754}
  (\bibinfo {year} {2022})}\BibitemShut {NoStop}%
\bibitem [{\citenamefont {Zhang}\ \emph {et~al.}(2018)\citenamefont {Zhang},
  \citenamefont {Zhao}, \citenamefont {Zhou}, \citenamefont {Xue},
  \citenamefont {Ma},\ and\ \citenamefont {Yang}}]{zhang2018strong}%
  \BibitemOpen
  \bibfield  {author} {\bibinfo {author} {\bibfnamefont {J.}~\bibnamefont
  {Zhang}}, \bibinfo {author} {\bibfnamefont {B.}~\bibnamefont {Zhao}},
  \bibinfo {author} {\bibfnamefont {T.}~\bibnamefont {Zhou}}, \bibinfo {author}
  {\bibfnamefont {Y.}~\bibnamefont {Xue}}, \bibinfo {author} {\bibfnamefont
  {C.}~\bibnamefont {Ma}},\ and\ \bibinfo {author} {\bibfnamefont
  {Z.}~\bibnamefont {Yang}},\ }\bibfield  {title} {\bibinfo {title} {Strong
  magnetization and chern insulators in compressed graphene/cri 3 van der waals
  heterostructures},\ }\href@noop {} {\bibfield  {journal} {\bibinfo  {journal}
  {Physical Review B}\ }\textbf {\bibinfo {volume} {97}},\ \bibinfo {pages}
  {085401} (\bibinfo {year} {2018})}\BibitemShut {NoStop}%
\bibitem [{pri()}]{private_tomoya}%
  \BibitemOpen
  \href@noop {} {}\bibinfo {howpublished} {Private communication with Tomoya
  Asaba}\BibitemShut {NoStop}%
\bibitem [{\citenamefont {Hillebrecht}\ \emph {et~al.}(2004)\citenamefont
  {Hillebrecht}, \citenamefont {Ludwig},\ and\ \citenamefont
  {Thiele}}]{hillebrecht2004_about}%
  \BibitemOpen
  \bibfield  {author} {\bibinfo {author} {\bibfnamefont {H.}~\bibnamefont
  {Hillebrecht}}, \bibinfo {author} {\bibfnamefont {T.}~\bibnamefont
  {Ludwig}},\ and\ \bibinfo {author} {\bibfnamefont {G.}~\bibnamefont
  {Thiele}},\ }\bibfield  {title} {\bibinfo {title} {About trihalides with tii3
  chain structure: Proof of pair forming of cations in $\beta$-rucl3 and rubr3
  by temperature dependent single crystal x-ray analyses},\ }\href@noop {}
  {\bibfield  {journal} {\bibinfo  {journal} {Zeitschrift für anorganische und
  allgemeine Chemie}\ }\textbf {\bibinfo {volume} {630}},\ \bibinfo {pages}
  {2199} (\bibinfo {year} {2004})}\BibitemShut {NoStop}%
\bibitem [{\citenamefont {Zhou}\ \emph {et~al.}(2013)\citenamefont {Zhou},
  \citenamefont {Zhang},\ and\ \citenamefont {Deng}}]{zhou2013_3n}%
  \BibitemOpen
  \bibfield  {author} {\bibinfo {author} {\bibfnamefont {Y.-C.}\ \bibnamefont
  {Zhou}}, \bibinfo {author} {\bibfnamefont {H.-L.}\ \bibnamefont {Zhang}},\
  and\ \bibinfo {author} {\bibfnamefont {W.-Q.}\ \bibnamefont {Deng}},\
  }\bibfield  {title} {\bibinfo {title} {A 3n rule for the electronic
  properties of doped graphene},\ }\href@noop {} {\bibfield  {journal}
  {\bibinfo  {journal} {Nanotechnology}\ }\textbf {\bibinfo {volume} {24}},\
  \bibinfo {pages} {225705} (\bibinfo {year} {2013})}\BibitemShut {NoStop}%
\bibitem [{\citenamefont {Crippa}\ \emph {et~al.}(2024)\citenamefont {Crippa},
  \citenamefont {Bae}, \citenamefont {Wunderlich}, \citenamefont {Mazin},
  \citenamefont {Yan}, \citenamefont {Sangiovanni}, \citenamefont {Wehling},\
  and\ \citenamefont {Valent{\'\i}}}]{crippa2024heavy}%
  \BibitemOpen
  \bibfield  {author} {\bibinfo {author} {\bibfnamefont {L.}~\bibnamefont
  {Crippa}}, \bibinfo {author} {\bibfnamefont {H.}~\bibnamefont {Bae}},
  \bibinfo {author} {\bibfnamefont {P.}~\bibnamefont {Wunderlich}}, \bibinfo
  {author} {\bibfnamefont {I.~I.}\ \bibnamefont {Mazin}}, \bibinfo {author}
  {\bibfnamefont {B.}~\bibnamefont {Yan}}, \bibinfo {author} {\bibfnamefont
  {G.}~\bibnamefont {Sangiovanni}}, \bibinfo {author} {\bibfnamefont
  {T.}~\bibnamefont {Wehling}},\ and\ \bibinfo {author} {\bibfnamefont
  {R.}~\bibnamefont {Valent{\'\i}}},\ }\bibfield  {title} {\bibinfo {title}
  {Heavy fermions vs doped mott physics in heterogeneous ta-dichalcogenide
  bilayers},\ }\href@noop {} {\bibfield  {journal} {\bibinfo  {journal} {Nature
  Communications}\ }\textbf {\bibinfo {volume} {15}},\ \bibinfo {pages} {1357}
  (\bibinfo {year} {2024})}\BibitemShut {NoStop}%
\end{thebibliography}%
\end{document}